\documentclass[twoside, twocolumn]{article}
\usepackage[utf8]{inputenc}
\usepackage[T1]{fontenc}
\usepackage{graphicx}
\usepackage[switch]{lineno}
\usepackage{longtable}
\usepackage{wrapfig}
\usepackage{rotating}
\usepackage{multicol}
\usepackage{placeins}  % provides \FloatBarrier
\usepackage[normalem]{ulem}
\usepackage{makecell}
\usepackage{newunicodechar}
\newunicodechar{Σ}{$\Sigma$}
\usepackage{amsmath}
\usepackage{cuted}   % for the strip environment
\usepackage{xcolor} % for \textcolor
\usepackage{caption} % for \captionof
\usepackage{amssymb}
\usepackage{float}
\usepackage{enumitem}
\usepackage{booktabs} % For prettier tables
\usepackage{capt-of}
\usepackage{hyperref}
\usepackage[acronym,nogroupskip,nonumberlist]{glossaries}
\usepackage[stylemods]{glossaries-extra}
\makeglossaries
\usepackage{abstract}%% easy abstract environment (from frontmatter pkg "ltxfront")
\usepackage[export]{adjustbox}%% expanded control over image, minipages, etc
\usepackage{sectsty}%% control of section heading appearance in default and script doc classes
\usepackage{setspace}%% implements \setstretch for line spacing
\usepackage{amsthm}%% formal mathematics environments
\usepackage{mathtools}%% bugfixing and additional tools for amsmath
\usepackage{lastpage}%% allows ref to lastpage
\usepackage{amsfonts}%% formal math fonts
\usepackage{mathptmx}%% ghostscript/postscript fonts and font loading options
\usepackage{charter}%% ghostscript/postscript fonts outside of math envs
\usepackage{times}%% more ghostscript/postscript
\usepackage{bm}%% bold math
\usepackage[%% full (expands on capt-of) control over appearance of float captions
format=plain,
justification=justified,
singlelinecheck=false,
font={small, sf},
labelfont=bf,
labelsep=space
]{caption}%% global preamble, use \captionsetup{} to config env)

\usepackage[compact]{titlesec}%% replacement of sectioning macros, coexsiting with old stuff
\usepackage{fancyhdr}%% improved headers and footers
\usepackage{fnpos}%% footnote positioning tools
\usepackage{sidecap}%% control of figure and caption positioning and margin spill
\usepackage{gensymb}%% inter-environment consistent measurment unit symbols
\usepackage{upgreek}%% easy lower and uppercase nonitalicized greek letters
\usepackage{soul}%% spaceout and underline macros

\usepackage{array}%% modernization of array and tabular envs
\usepackage{droidsans}%% use android sanserif fonts...

\usepackage[usenames,dvipsnames]{xcolor}%% text color macros
\usepackage[version=3]{mhchem}%% for writing chemical formulae
\usepackage[super,sort&compress,comma]{natbib}%% prefered citation engine
\usepackage[%%rsc page standard follows
left=1.5cm,
right=1.5cm,
top=1.785cm,
bottom=2.0cm
]{geometry}

\definecolor{cream}{RGB}{222,217,201}%% required for use of rsc frontmatter

\date{\today}
\title{Using Machine Learning to Explore Defect Configurations in Cd/Zn-Se/Te Compounds }
\hypersetup{
 pdfauthor={Md. Habibur Rahman},
 pdftitle={DeFecT‑FF: Accelerated Modeling of Defects in Cd/Zn-Te/Se/S Compounds Combining High-Throughput DFT and Machine Learning Force Fields},
 pdfkeywords={},
 pdfsubject={},
 pdfcreator={Emacs 29.0.50 (Org mode 9.5.3)}, 
 pdflang={English}}
\begin{document}

\pagestyle{fancy}
\thispagestyle{plain}
\fancypagestyle{plain}{
\renewcommand{\headrulewidth}{0pt}
}

\makeFNbottom
\makeatletter
\renewcommand\LARGE{\@setfontsize\LARGE{15pt}{17}}
\renewcommand\Large{\@setfontsize\Large{12pt}{14}}
\renewcommand\large{\@setfontsize\large{10pt}{12}}
\renewcommand\footnotesize{\@setfontsize\footnotesize{7pt}{10}}
\makeatother

\renewcommand{\thefootnote}{\fnsymbol{footnote}}
\renewcommand\footnoterule{\vspace*{1pt}% 
\color{cream}\hrule width 3.5in height 0.4pt \color{black}\vspace*{5pt}} 
\setcounter{secnumdepth}{5}

\makeatletter 
\renewcommand\@biblabel[1]{#1}            
\renewcommand\@makefntext[1]{\noindent\makebox[0pt][r]{\@thefnmark\,}#1}
\makeatother 
\renewcommand{\figurename}{\small{Figure}~}
\sectionfont{\sffamily\Large}
\subsectionfont{\normalsize}
\subsubsectionfont{\bf}
\setstretch{1.125}
\setlength{\skip\footins}{0.8cm}
\setlength{\footnotesep}{0.25cm}
\setlength{\jot}{10pt}
\titlespacing*{\section}{0pt}{4pt}{4pt}
\titlespacing*{\subsection}{0pt}{15pt}{1pt}

\fancyfoot{}
\fancyfoot[RO]{\footnotesize{\sffamily{1--\pageref{LastPage} ~\textbar  \hspace{2pt}\thepage}}}
\fancyfoot[LE]{\footnotesize{\sffamily{\thepage~\textbar\hspace{3.45cm} 1--\pageref{LastPage}}}}
\fancyhead{}
\renewcommand{\headrulewidth}{0pt} 
\renewcommand{\footrulewidth}{0pt}
\setlength{\arrayrulewidth}{1pt}
\setlength{\columnsep}{6.5mm}
\setlength\bibsep{1pt}

\makeatletter 
\newlength{\figrulesep} 
\setlength{\figrulesep}{0.5\textfloatsep} 

\newcommand{\topfigrule}{\vspace*{-1pt}% 
\noindent{\color{cream}\rule[-\figrulesep]{\columnwidth}{1.5pt}} }

\newcommand{\botfigrule}{\vspace*{-2pt}% 
\noindent{\color{cream}\rule[\figrulesep]{\columnwidth}{1.5pt}} }

\newcommand{\dblfigrule}{\vspace*{-1pt}% 
\noindent{\color{cream}\rule[-\figrulesep]{\textwidth}{1.5pt}} }

\makeatother
\twocolumn[
\begin{@twocolumnfalse}
\vspace{1em}
\sffamily
\begin{tabular}{m{4.5cm} p{13.5cm} }
& \noindent\LARGE{\textbf{DeFecT-FF: A Machine Learning Force Field Framework for High Throughput Defect Modeling in CdTe-Based Solar Cells}}\\%

\vspace{0.3cm} & \vspace{0.3cm} \\
& \noindent\large{Md Habibur Rahman\textsuperscript{a}, Maitreyo Biswas\textsuperscript{a} and Arun Mannodi-Kanakkithodi\textsuperscript{a}}\\

\end{tabular}

\begin{abstract}

We developed a framework for predicting the energies and ground state configurations of native point defects, extrinsic dopants and impurities, and defect complexes across zincblende-phase Cd/Zn-Te/Se/S compounds, important for CdTe-based solar cells. This framework, named \texttt{DeFecT-FF}, is powered by high-throughput density functional theory (DFT) computations and crystal graph-based machine learning force field (MLFF) models trained on the DFT data. The Cd/Zn-Te/Se/S chemical space is chosen because alloying at Cd or Te sites is a promising avenue to tailor the electronic and defect properties of the CdTe absorber layer to potentially improve solar cell performance. The sheer number of defect configurations achievable when considering all possible singular defects and their combinations, symmetry-breaking operations, and defect charge states, as well as the expense of running large supercell calculations, makes this an ideal problem for developing accurate and widely-applicable force field models. Here, we introduce our dataset of structures and energies from HSE06 geometry optimization, including bulk and alloyed supercells with and without defects. Data were gradually expanded using active learning and accurate MLFF models were trained to predict energies and atomic forces across different charge states. Via accelerated prediction and screening, we identified many new low energy defect configurations and obtained high-fidelity defect formation energy diagrams using HSE06 calculations with spin-orbit coupling. The \texttt{DeFecT-FF} framework has been released publicly as an online tool on the nanoHUB platform, allowing users to upload any crystallographic information file, generate defects of interest, and compute defect formation energies as a function of Fermi level and chemical potential conditions, thus bypassing expensive DFT calculations. 

\end{abstract}
\end{@twocolumnfalse}
\vspace{0.6cm}
]

\renewcommand*\rmdefault{bch}\normalfont\upshape
\rmfamily
\section*{}
\vspace{-1cm}

\footnotetext{%
  \textsuperscript{a}School of Materials Engineering, Purdue University, West Lafayette, IN 47907, USA; E-mail: amannodi@purdue.edu
  
}

\section{Introduction}

Advancements in solar cell technologies are crucial for meeting global energy demands and supporting the transition to a decarbonized grid \cite{Srisuda1, Srisuda2, a2bcx4_abx2, 11133092}. Among photovoltaic (PV) technologies, CdTe ranks second after crystalline Si, accounting for about 7\% of the global market \cite{CdTe_1, Srisuda1, Srisuda2}. Its commercial success arises from a direct bandgap of $\sim$1.5 eV, high absorption coefficient ($>5\times10^{5}$ cm$^{-1}$) \cite{CdTe_2, eduardo_men_ndez_proupin_dae1679b, fadhil_k__alfadhili_1e77f8db, o__de_melo_1bf7fe05, kaiying_luo_60e9a185, wei_chao_chen_652bd27c, kai_shen_86d534d1, jinshui_miao_634f66d0, a_gayt_n_garc_a_23ea61bf, xin_zheng_02d025a7}, low production cost, and good thin-film conductivity \cite{CdTe_3}. However, its maximum efficiency of 22.3\% remains below the $\sim$30\% theoretical limit, mainly due to Shockley-Read-Hall (SRH) recombination associated with grain boundaries, point defects, and dislocations \cite{Freysoldt2014-en}. Native defects and impurities create trap states within the bandgap that act as nonradiative recombination centers \cite{Ganose_Scanlon_2022, Mannodi-Kanakkithodi_2023}, such as Cd vacancies (V$_{\mathrm{Cd}}$), which accelerate carrier recombination and can reduce power conversion efficiency by nearly 5\% \cite{sean_1, Turiansky, aneta_wardak_ab3492d5, p__d__hatton_26561f92, michael_a__scarpulla_043ff246}. \\

CdTe often suffers from low hole density, limiting its PV efficiency \cite{Srisuda1, Srisuda2}. Cu is commonly introduced as an acceptor dopant via high-temperature CdCl$_2$ annealing, where Cu and Cl diffuse at 10$^{17}$–10$^{19}$ cm$^{-3}$, altering electronic properties \cite{Yang_Yin, Srisuda1, Srisuda2}. While Cu$_{i}$ and Cl$_{Te}$ act as shallow donors and Cu$_{Cd}$ as a non-shallow acceptor forming complexes such as (Cu$_{i}$+Cu$_{Cd}$) and (Cl$_{i}$+Cu$_{Cd}$)$^{2+}$ \cite{Krasikov_1, Krasikov_2, CdTe_1}, Cu doping typically yields suboptimal hole density ($\sim$10$^{14}$ cm$^{-3}$) compared to the ideal value of $\sim$10$^{16}$ cm$^{-3}$ \cite{CdTe_3}. In contrast, group V dopants such as As achieve higher hole densities without reducing carrier lifetime \cite{Ablekim, CdTe_1}. Se alloying to create CdSe$_{x}$Te$_{1-x}$ further enhances PV efficiency by improving absorption, band alignment, and carrier lifetimes \cite{Ablekim, Fiducia2019-wa}. ZnTe, with favorable band alignment, serves as an efficient hole transport layer \cite{Gorai_Krasikov}. Thus, exploring the defect chemistry of Cd/Zn–S/Se/Te alloys is vital to improving CdTe- and CdSeTe-based thin-film solar cells \cite{De_Souza_Harrington_2023}. \\

In semiconductors, point defects can trap or release electrons, and thus they tend to exist in multiple charge states $q$ depending on the Fermi level position ($E_F$) within the band gap. For each defect, the charge transition level $\varepsilon(q/q')$ marks the $E_F$ value at which the defect switches from charge state $q$ to $q'$; below this level one charge state is preferred, and above it the other. These deep or shallow nature of these transition levels determine whether a defect acts as an electron donor, acceptor, or recombination center, and are central to understanding semiconductor doping and device performance. Defect levels are experimentally measured using cathodoluminescence, photoluminescence, optical spectroscopy, or deep-level transient spectroscopy (DLTS) \cite{Lang2003}. These methods face significant challenges in sample preparation and assigning measured levels to specific defects \cite{Wickramaratne, Kim_Gelczuk}. To overcome this, density functional theory (DFT) is widely used to calculate defect formation energy (E$^{f}$) as a function of E$_{F}$, defect charge state (\textit{q}), and chemical potential ($\mu$) \cite{Freysoldt2014-en, Broberg_Bystrom, gw_def, Lee_Din, Grill2004-pk, Buckeridge_2019}. DFT enables identification of donor- and acceptor-type defects, shallow or deep defect levels, type of equilibrium conductivity, defect concentrations, and carrier capture rates \cite{Mannodi-Kanakkithodi2022-ck, Turiansky, Defect_Energetics_JPCC, Kim_Park_Hood_Walsh_2019, Lyons2017-sp}. When an appropriate level of theory is applied, DFT-computed charge transition levels compare well with experiments \cite{Mannodi-Kanakkithodi2022-ck, Kim_Gelczuk, snb_1, snb_2}. However, DFT is computationally expensive and scales poorly with system size, making it difficult to explore the vast configurational space of vacancies, interstitials, antisites, and defect complexes across many compounds and charge states \cite{mba, Mannodi-HP1}. \\

The prediction of defect properties can be accelerated by integrating DFT simulations with machine learning (ML) approaches such as crystal graph neural networks (GNNs) \cite{CGCNN, Choudhary, Chen_Li_Bruna_2017}. GNNs effectively represent and predict the energies and properties of molecules, polymers, and crystalline materials \cite{Kipf_Welling_2016, Witman_2023} by transforming atomic structures into graphs where atoms are nodes and bonds are edges \cite{m3gnet}. They learn intricate structural representations to predict properties such as formation or decomposition energy, bandgap, and defect formation energy while reducing computational cost. In prior work, we used GNNs to predict and screen native defects and functional impurities in group IV, III–V, and II–VI zincblende semiconductors \cite{Rahman_Gollapalli}, covering vacancies, interstitials, anti-site, and extrinsic defects. While the models predicted charge-dependent defect formation energies for different chemical potential conditions, several limitations were observed: (1) training on a broad chemical space (34 compounds) led to large errors for specific compositions; (2) models showed good performance on binaries but lower accuracy for alloy systems such as CdSe$_{x}$Te$_{1-x}$ and Cd$_{x}$Zn$_{1-x}$Te \cite{Rahman_Rojsatien, Yang_Yin, Krasikov_2, Srisuda2, Defect_Energetics_JPCC, Rahman_Mannodi-Kanakkithodi_2025b, Lany_soc}; (3) reliance on modest 64-atom \(2\times2\times2\) supercells limited defect complex modeling; (4) use of the semi-local GGA-PBE functional (GGA: Generalized Gradient Approximation, PBE: Perdew–Burke–Ernzerhof) inherently limited prediction fidelity \cite{GGA_gaps, HP_hse, Lyons_Van}; and (5) dependence on gradient-free optimization with GNN models prevented more efficient gradient-based geometry optimization. \\

To overcome prior limitations from casting too wide a chemical space, using smaller supercells, and relying on semi-local functionals, we developed a more comprehensive, multi-fidelity methodology. Our approach begins with an initial dataset of bulk and defect configurations spanning Cd/Zn–Te/Se/S binary and multi-nary compounds. These structures were first computed using the PBE functional, providing a baseline set of bulk and defect configurations and their energies, with a substantial portion of the PBE dataset compiled from our previously published works \cite{Mannodi-ZB, Mannodi-Kanakkithodi2022-ck}. Initial GNN models trained on the PBE dataset ~\cite{Bapst_Keck} served as the foundation for predicting defect properties over a pre-defined defect chemical space containing thousands of vacancies, interstitials, antisites, substitutional defects, and defect complexes. Defect enumeration included all relevant native defects as well as group-V dopants (N, P, As, Sb, Bi) which are promising for achieving p-type conductivity and unintentional impurities such as Cl and O which are known to strongly influence the performance of CdTe and CdSe$_x$Te$_{1-x}$ solar cells~\cite{Li2017-ez, Corsin_2016}. The Cd/Zn–Te/Se/S chemical space was chosen due to its relevance to Se grading, Cd–Zn interfaces, and absorber composition tuning in CdTe solar cells, where exploring all low-energy native and extrinsic defects across these compositions provides a comprehensive dataset for experimental comparison. Although ZnS and ZnSe are not primary absorbers, they remain chemically informative, while CdS functions as an important buffer layer \cite{CdTe_1, CdTe_2, CdTe_3}. \\

\begin{figure*}[t]
\centering
\includegraphics[width=.9\linewidth]{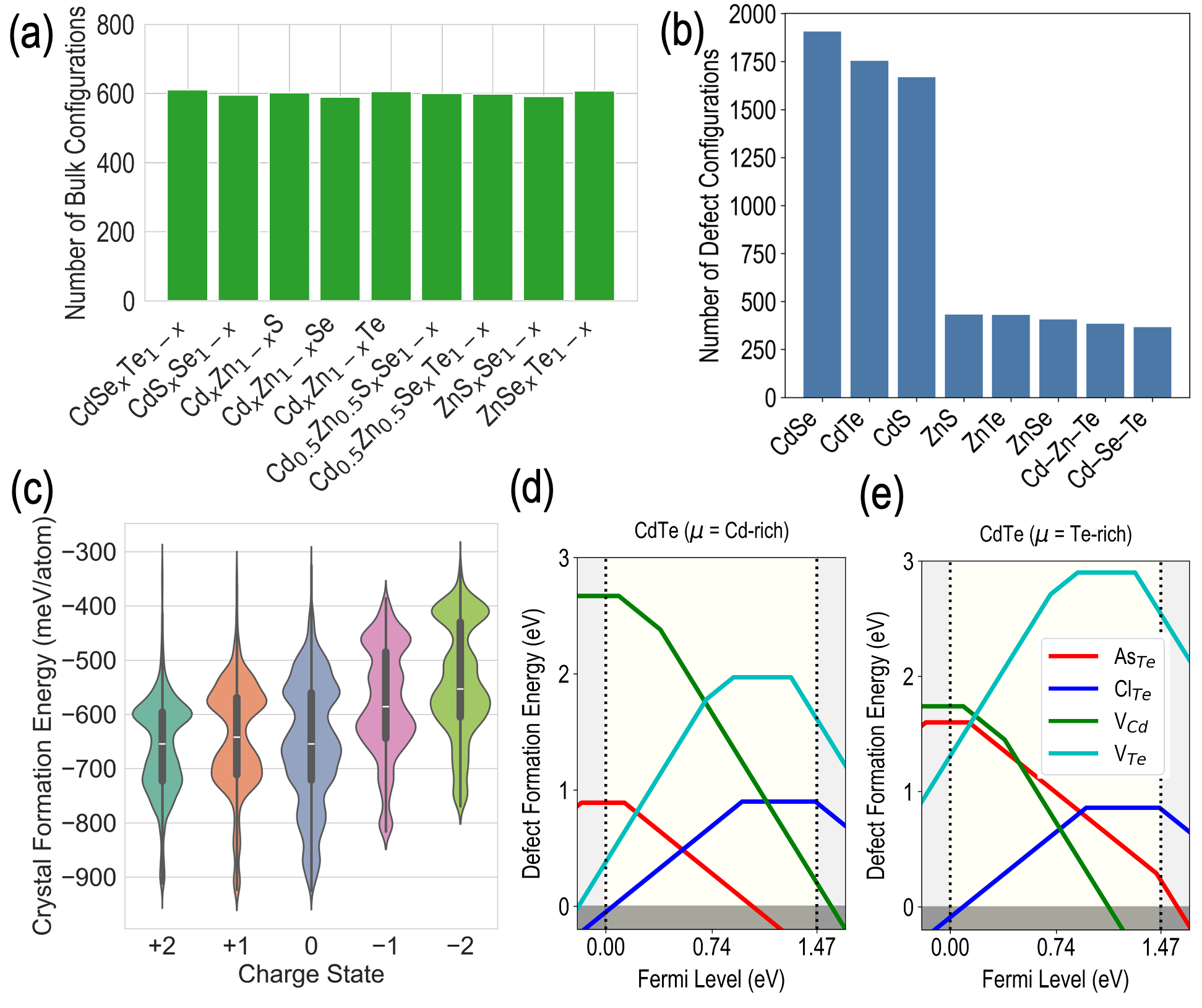}
\caption{\label{fig:hse_data}
Statistics of the HSE06 dataset: (a) Number of bulk configurations from the CdSe$_{x}$Te$_{1-x}$, CdS$_{x}$Se$_{1-x}$, Cd$_{x}$Zn$_{1-x}$S, Cd$_{x}$Zn$_{1-x}$Se, Cd$_{x}$Zn$_{1-x}$Te, Cd$_{0.5}$Zn$_{0.5}$S$_{x}$Se$_{1-x}$, Cd$_{0.5}$Zn$_{0.5}$Se$_{x}$Te$_{1-x}$, ZnS$_{x}$Se$_{1-x}$, and ZnSe$_{x}$Te$_{1-x}$ compositions. (b) Distribution of defect configurations across the Cd–chalcogen and Zn–chalcogen binaries and ternaries. (c) Violin plots of crystal formation energies (meV/atom) for the entire dataset across five charge states ($+2$ to $-2$).  (d,e) Defect formation energy diagrams for CdTe under Cd-rich and Te-rich conditions from HSE06 functional, highlighting the relative stability of key native (V$_{Cd}$, V$_{Te}$) and extrinsic defects (As$_{Te}$, Cl$_{Te}$).}
\end{figure*}

To improve prediction accuracy, active learning was employed to iteratively generate new DFT data and refine the GNN models \cite{AL_1, AL_2, AL_3, AL_4, AL_5, Gentle_2010, Farache_Verduzco}. Following convergence of the active learning scheme, we performed higher-fidelity HSE06 (Heyd–Scuseria–Ernzerhof) ~\cite{hse} calculations on a curated subset of representative PBE-relaxed structures to obtain more accurate bandgaps, charge transition levels, and defect formation energies. GNN models, specifically using the M3GNet \cite{m3gnet} architecture for machine learning force fields (MLFFs), were then trained on the HSE06 data and subsequently used for new predictions. Our complete methodology is summarized as:  
PBE data collection $\rightarrow$ initial PBE GNN model training $\rightarrow$ active-learning–driven expansion of the PBE dataset $\rightarrow$ HSE06 refinement of a subset of the PBE dataset $\rightarrow$ training MLFF models at HSE06 accuracy. The next few sections describe our methodology and results in detail, highlighting the following major contributions of this work: 

\vspace{0.3cm}
\noindent 
\begin{itemize}
    \item Construction of the largest unified HSE06 defect dataset across Cd/Zn–Te/Se/S compositions, including native and extrinsic defects, and defect complexes, simulated in five charge states.
    \item Development of an HSE06 MLFF-based defect optimization workflow that is orders of magnitude faster than full DFT. 
    \item Release of \texttt{DeFecT-FF}, an online nanoHUB tool \cite{defect-ff} with the following workflow: input bulk structure + list of defect candidates $\rightarrow$ generation of defect structures with symmetry breaking $\rightarrow$ MLFF optimization across five charge states $\rightarrow$ selection of the lowest-energy configurations $\rightarrow$ final HSE06+SOC (spin–orbit coupling) calculation. \\
\end{itemize}

\section{Description of the DFT Datasets}

The entire Cd/Zn–Te/Se defect chemical space is extraordinarily large, as summarized in the Supporting Information (SI), \textbf{Table~\ref{tab:defect_counts_v1}}. Treating every symmetry-inequivalent site in a $3\times3\times3$ (216-atom) cubic zincblende supercell results in thousands of possible defect configurations per composition, driven by the combinatorial explosion of native vacancies, self-interstitials, anti-site substitutions, eight types of extrinsic defects (dopants—Cu, As, P, N, Sb, Bi; impurities—Cl, O), and all unordered defect complexes, as illustrated in \textbf{Figure~\ref{fig:chemical_sapce}}. Even when restricted to only neutral charge states, the number of required DFT calculations grows to nearly $\approx$ 0.9 million for the entire chemical space. Thankfully, the defect structures collected from published literature and our prior work~\cite{Mannodi-ZB, Mannodi_Toriyama_2020, Mannodi-Kanakkithodi2022-ck, Rahman_Gollapalli} already represent a physically meaningful and chemically rich subset of this much larger defect chemical space. These literature-curated PBE-optimized structures span multiple compositions, charge states, and defect chemistries, providing a robust foundation for training the initial GNN models. We then employed an active-learning workflow to selectively launch new PBE calculations in regions of the chemical space where the ML model exhibited high uncertainty or sparse representation. Details of this dataset-building strategy are provided in the SI. \\

A carefully selected subset of PBE-optimized structures was further used to perform hybrid HSE06 calculations after adjusting the defect supercell lattice parameters to values from HSE06 optimization, before volume-fixed relaxation, as summarized in the \textbf{Table~\ref{tab:lattice_parameters}}. The resulting HSE06 dataset captures more than half of the structural and chemical diversity present in the full PBE dataset, and the statistical distribution of this HSE subset  is illustrated in \textbf{Figure~\ref{fig:hse_data}}. Importantly, all HSE calculations preserved every ionic-relaxation snapshot—not only the final minimum-energy configuration—yielding thousands of intermediate structures with their corresponding energies, forces, and stresses. This provides a significantly enriched dataset that captures full relaxation pathways rather than ground-state endpoints alone. For a complete statistical overview of the PBE and HSE datasets across all charge states, readers are referred to \textbf{Table~\ref{table:bulk_compounds_list}}, \textbf{Table~\ref{table:initial_pbe_dataset}} and \textbf{Table~\ref{table:hse_dataset}}. Additional dataset visualization is provided in \textbf{Figure~\ref{fig:pbe-violins}}, \textbf{Figure~\ref{fig:initial_pbe_dataset}}, and \textbf{Figure~\ref{fig:hse-violins}}. In summary, our data generation pipeline proceeds as follows: literature data collection $\rightarrow$ initial PBE GNN model training $\rightarrow$ active-learning–driven expansion of the PBE dataset $\rightarrow$ HSE06 refinement of a subset of the PBE dataset. \\

\begin{figure*}[t]
\centering
\includegraphics[width=.9\linewidth]{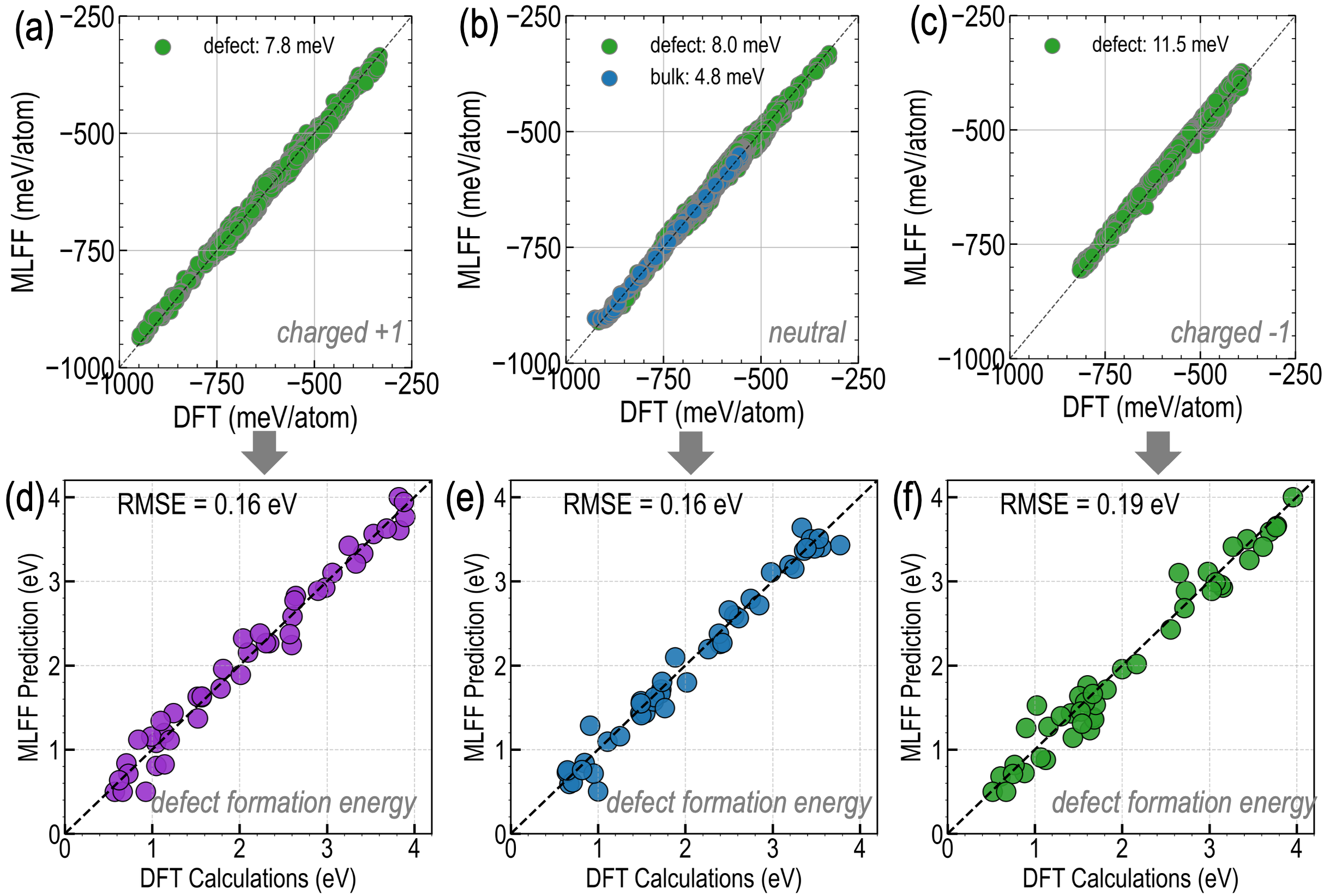}
\caption{\label{fig:hse_mlff_models}
(a–c) Parity plots comparing crystal formation energies from DFT and MLFF predictions for three representative charge states: (a) q = +1, (b) q = 0 (neutral), and (c) q = –1. The MLFF accurately reproduces the DFT energies with small errors, as indicated by the RMSE values shown in each panel. (d–f) Defect formation energies under Cd-rich condition computed using MLFF predictions for a subset of the defect configurations shown in panels (a–c), compared against values from full DFT. The MLFF defect energies were obtained by adding DFT reference energies and applying charge corrections to the MLFF-predicted total energies.}
\end{figure*}

\begin{figure*}[t]
  \centering
  \includegraphics[width=\textwidth]{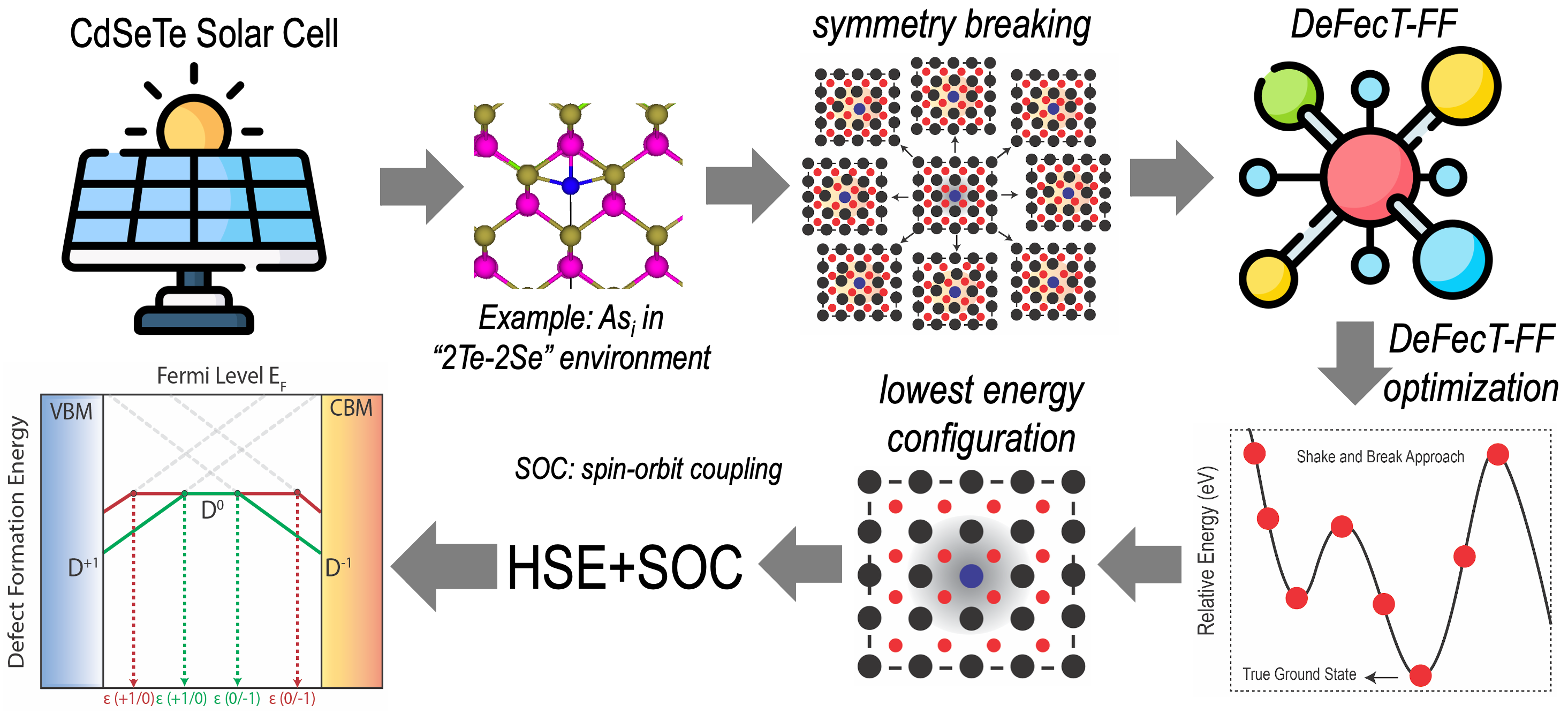}%  <-- changed
  \caption{Workflow for accelerated defect predictions using the \texttt{DeFecT-FF} framework. An initial defect structure (example: As$_i$ in a mixed ``2Te--2Se'' local environment) is constructed and passed through the ShakeNBreak \cite{snb_1} symmetry-breaking procedure to generate a diverse set of competing defect geometries. These distorted configurations are rapidly relaxed using the rigorously optimized machine-learned force field to identify the lowest-energy structure prior to high-fidelity DFT calculations. The optimized geometry is then used to perform static HSE+SOC calculation, yielding accurate defect formation energy diagrams.}
\label{fig:workflow}
\end{figure*}

DFT calculations were performed using VASP~\cite{vasp} on the Negishi cluster at Purdue University, utilizing nodes equipped with two AMD EPYC 7763 ``Milan'' CPUs @ 2.2 GHz (128 cores per node, 256 GB memory). HSE06 relaxation jobs were run using 512 cores (4 nodes). The MLFF relaxations were performed on a single compute node with 16 cores. Under these conditions, a single HSE06 geometry optimization for a charged defect in a 216-atom supercell requires approximately 4,096 core-hours, whereas the DeFecT-FF relaxation completes in approximately 0.5 core-hours, representing a speedup exceeding four orders of magnitude. \\

\section{MLFF models trained at hybrid functional accuracy}

We initially trained models on the PBE dataset using the ALIGNN framework \cite{ALIGNN} and then employed an active learning strategy to launch new DFT calculations by targeting regions of the chemical space with largest prediction uncertainty. We then refined a representative subset of the PBE-optimized structures using the HSE06 functional and used these data to train an M3GNet-based \cite{m3gnet} MLFF, using DFT-derived configurations, energies, forces, and stresses. Readers are referred to the SI for additional details on the ALIGNN training procedure and the active learning workflow used in this work. These discussions are supplemented by a series of figures in the SI (\textbf{\hyperref[fig:al_workflow]{Figures~S5–S13}}), which collectively illustrate the active learning pipeline, model transferability, ALIGNN-based structural optimization, comparisons with other MLFF models, and MLFF models on PBE dataset. The active learning workflow~\cite{Farache_Verduzco} employs an ensemble of ALIGNN models ~\cite{ALIGNN}, with the standard deviation of their energy predictions serving as the uncertainty metric. In each active learning batch, the top 200 structures with the highest prediction uncertainty (generally 5--10 meV/atom standard deviation) are selected for additional DFT calculations. Convergence is assessed by monitoring the fraction of newly queried structures falling within the model's confidence interval and the saturation of test set RMSE. After 3--4 iterations, both metrics indicated convergence. Additional details on the active learning pipeline, including sensitivity analysis, are provided in the Supporting Information (Figures S5--S13). \\

HSE06 calculations were performed using \(\Gamma\)-point only \cite{Mosquera2}, with a reduced plane-wave energy cutoff of 400 eV. The convergence thresholds for geometry optimization were set to \(10^{-6}\)~eV for energy and 0.01 eV \AA$^{-1}$ for forces. \textbf{Figure~\ref{fig:hse_data}} shows the statistics of the compiled HSE06 dataset in terms of the number of bulk Cd/Zn-Te/Se/S composition structures, and different types of defects. The HSE06 dataset represents 53.4\% of the GGA dataset for bulk ($q=0$) structures, and 63.8\%, 71.6\%, 79.5\%, 72.4\%, and 65.8\% for defect structures in charge states $q=+2$, $q=+1$, $q=0$, $q=-1$, and $q=-2$, respectively. Despite the smaller size, it remains well representative of the defect types and structural diversity of the entire chemical space. Violin plots showing the spread of the crystal formation energy (CFE, a per-atom energy difference between the crystal and its constituent atoms) values in the HSE06 dataset are presented in \textbf{Figure~\ref{fig:hse-violins}}. CFE is defined as the per-atom energy difference between the total energy of the crystal and the sum of elemental reference energies of its constituent elements, i.e., CFE $= (E_{\text{crystal}} - \sum_i n_i E_{\text{ref},i}) / N$, where $E_{\text{ref},i}$ is the reference energy of element $i$ and $N$ is the total number of atoms in the supercell. The reference states for all elements are taken as the lowest-energy phases from the Materials Project~\cite{Jain_Ong}.

Several MLFF frameworks have been developed for materials property prediction, including pretrained universal models such as MACE~\cite{mace}, CHGNet~\cite{chgnet}, and M3GNet~\cite{m3gnet}. However, these models are trained predominantly on neutral bulk structures from databases such as the Materials Project~\cite{Jain_Ong}, and consequently generalize poorly to charged defect configurations in semiconductors. DeFecT-FF addresses this gap through four key innovations: (i) charge-state-resolved models that explicitly capture the structural and energetic signatures of defects in five charge states; (ii) a multi-fidelity active learning pipeline that efficiently bridges PBE and HSE06 levels of theory; (iii) training on defect-specific data including intermediate relaxation snapshots, symmetry-broken geometries from ShakeNBreak~\cite{snb_2}, and defect complexes; and (iv) deployment as an end-to-end nanoHUB tool for community use. \\

Before training the MLFF models, we first evaluated the performance of state-of-the-art pretrained force-field frameworks, namely MACE \cite{mace}, CHGNet \cite{chgnet} and M3GNet \cite{m3gnet}, by applying them directly to our PBE dataset. Although these models have demonstrated strong predictive capabilities on their training domains, they generalized poorly to the chemically diverse Cd/Zn–Te/Se/S systems investigated in this work as summarized in \textbf{Table~\ref{tab:pretrained_mlff_comparison}}. Root mean square error (RMSE) in CFE prediction ranged from 60 to 100 meV/atom, which it will turn out are much larger than errors from fine-tuned models. These shortcomings indicate that the pretrained models lack sufficient exposure to the chalcogenide defect chemical space considered here. Consequently, this motivated the development of a dataset-specific M3GNet-based MLFF trained on DFT-derived configurations, energies, forces, and stresses, enabling the level of accuracy required for reliable modeling of defect thermodynamics and structural relaxations. \\

Parity plots for M3GNet-MLFF models trained on the HSE06 dataset are pictured in \textbf{Figure~\ref{fig:hse_mlff_models}(a--c)}, respectively for charge states \(q = +1\), \(q = 0\), and \(q = -1\). Models for the \(q=+2\) and \(q=-2\) charge states are presented in \textbf{Figure~\ref{fig:hse_m3gnet_model_old_2}}. Each parity plot compares the HSE06-computed CFE with MLFF-predicted values across different categories: bulk (pristine supercells without defects), and defects (bulk supercells containing a single defect or defect complex). Despite the reduced dataset size compared to the GGA dataset, the HSE-MLFF models achieve very good accuracy with low RMSE values across different structure types. The \(q = 0\) test set prediction RMSE ranges from 4.8 meV/atom for bulk structures to 7.8 meV/atom for defect structures. These errors are similar for \(q = +1\) and \(q = -1\) defect structures and remain below 12 meV/atom for all cases, which is quite reasonable given the range of CFE values in the dataset. We also simulated a limited number of CdTe dislocation core structures \cite{Fatih-CdTe-GB,Jinglong} and CdTe/ZnTe interface configurations with selected defects and included them in the training dataset. The model performance for these structures is shown in \textbf{Figure~\ref{fig:hse_m3gnet_model_old_1}} and \textbf{Figure~\ref{fig:hse_m3gnet_model_old_2}}. These results are not discussed in detail in the main text because the number of data points corresponding to dislocation cores and interfaces is relatively small. \\

The training dataset spans the entire Cd/Zn--Te/Se/S chemical space with all native defect types and extrinsic defect species across five charge states, covering 14 individual compounds that include 6 binaries and 8 ternary alloys. Table~\ref{tab:rmse_by_defect_type} provides a breakdown of model accuracy by defect type, showing consistent performance across vacancies (9.5 meV/atom), extrinsic substitutions (8.6 meV/atom), anti-site substitutions (8.1 meV/atom), and interstitials (7.6 meV/atom). Active learning specifically targeted underrepresented regions to ensure broad coverage of the defect chemical space. \\

\textbf{Figure~\ref{fig:single_vs_complex}} shows the MLFF predictions for neutral defect structures separately for single defects and defect complexes, revealing low RMSE values of 8.14~meV/atom and 9.23~meV/atom respectively. This suggests that the MLFF effectively captures both localized and collective defect relaxations, even for configurations with multiple defects. Even though MLFF prediction of crystal formation energy is highly accurate for all bulk and defect structures, a true evaluation of the prediction for defect configurations involves comparing defect formation energy (and defect level) predictions from MLFF and HSE06. We used the MLFF models to optimize a selected set of defects structures. For each defect, the MLFF was used to compute the total energy entering the standard defect formation energy expression:
\[
\Delta E_{f}(D^{q})
= E_{tot}(D^{q}) - E_{tot}(bulk)
+ \sum_{i} n_{i}\mu_{i}
+ q\bigl(E_{F} + E_{VBM}\bigr)
+ E_{corr}
\]
Here, $E_{tot}(D^{q})$ and $E_{tot}(bulk)$ are the total energies of the defect and pristine supercells respectively, $n_{i}$ and $\mu_{i}$ denote the stoichiometric changes and elemental chemical potentials, $q$ is the charge state, $E_{VBM}$ is the valence band maximum energy, $E_{F}$ is the Fermi level through the band gap, and $E_{corr}$ is the correction energy~\cite{Freysoldt2014-en} which accounts for spurious electrostatic interactions arising from the periodic repetition of charged defects and the compensating background charge in finite supercells. In this work, $E_{corr}$ is evaluated using the Freysoldt ~\cite{Freysoldt2014-en} charge correction scheme, which separates long-range Coulomb interactions from short-range defect-induced potentials and aligns the electrostatic potential between defect and bulk calculations~\cite{Freysoldt2014-en}. Additional methodological details and convergence tests for the charge correction are provided in the SI. Importantly, the reference energy $\mu_{i}$, $E_{VBM}$ and $E_{corr}$ were evaluated from DFT and added directly to the MLFF-derived bulk and defect energies. \\

\textbf{Figure~\ref{fig:hse_mlff_models}(d--f)} present parity plots for defect formation energy corresponding to $q=+1$ (at E$_{F}$=0), $q=0$, and $q=-1$ (at E$_{F}$=0). Across the entire validation set, the RMSE in defect formation energies obtained using MLFF-optimized geometries remains below $0.20$~eV, demonstrating that the MLFFs are sufficiently accurate in capturing both structural and energetic trends for charged and neutral defects. To further assess the influence of charge corrections on MLFF-derived defect energetics, we compared three approaches:  (i) using MLFF-predicted defect formation energies without any charge correction;  (ii) using MLFF defect formation energies corrected using a simple average offset of $0.20$~eV for $q=+2$, $0.10$~eV for $q=+1$, $0.10$~eV for $q=-1$, and $0.20$~eV for $q=+2$ defects; and (iii) adding known charge correction energies from DFT to MLFF-predicted defect formation energies (\textbf{Figure~\ref{fig:hse_mlff_models}(d--f)}). \\

The average charge-correction offsets ($\sim$0.20 eV for $|q|=2$, $\sim$0.10 eV for $|q|=1$) were derived empirically from extensive defect calculations across the Cd/Zn--Te/Se/S chemical space. Their near-uniformity arises from the similar supercell geometries and the narrow range of dielectric constants ($\sim$7--10) in this class of materials. We note that these offsets are specific to the present chemical space and may not be applicable to other chemistries with different dielectric properties, crystal structures, or supercell sizes. For applications beyond the Cd/Zn chalcogenide systems, explicit DFT-based Freysoldt corrections~\cite{freysoldt_new, Freysoldt2014-en} are necessary. \\

The comparison reveals that while uncorrected MLFF prediction shows errors close to 0.3 eV for all charge states, applying the average offset brings this error down closer to 0.2 eV which is similar to the error from adding known correction values, as listed in \textbf{Table~\ref{tab:rmse_charge_corrections}}. Parity plots comparing MLFF defect formation energy with DFT values for all three approaches are shown in \textbf{Figure~\ref{fig:charge_correction}}. \textbf{Figure~\ref{fig:transition_level_plot}} compares defect charge transition levels predicted by the MLFF with DFT reference values. Applying an average charge correction value leads to close agreement between MLFF- and DFT-predicted defect transition levels, with RMSE values of 0.25~eV, 0.23~eV, 0.22~eV, and 0.27~eV for the $\epsilon(+2/+1)$, $\epsilon(+1/0)$, $\epsilon(0/-1)$, and $\epsilon(-1/-2)$ transitions, respectively, whereas the DFT-based charge correction values are 0.22~eV, 0.21~eV, 0.20~eV, and 0.24~eV. The increase in RMSE from meV/atom (for CFE) to $\sim$ 0.2 eV (for defect formation energies) arises because the defect formation energy expression combines MLFF-predicted supercell energies with independently computed DFT-derived quantities (chemical potentials, $E_{\text{VBM}}$, charge corrections). The MLFF prediction errors for the defect and bulk supercells do not cancel perfectly due to distinct local atomic environments around the defect site. The residual $\sim$ 0.2 eV error reflects this imperfect cancellation combined with the reference-frame mismatch between MLFF and DFT contributions to the defect formation energy expression. This defect formation energy error is still reasonable and enormously useful for quick prediction and screening. \\

To assess the reliability of our MLFF models relative to full DFT, we randomly selected 100 representative bulk and defect configurations and relaxed each structure using both methods. The DFT- and MLFF-optimized geometries were then compared using SOAP descriptors \cite{dscribe_1, dscribe_2}, which provide a rotationally invariant fingerprint of the atomic environments. These high-dimensional descriptors were projected into a two-dimensional PCA \cite{PCA} (principle components analysis) space, allowing direct visualization of structural similarity. In \textbf{Figure~\ref{fig:pca}}, each DFT structure (blue) is paired with its corresponding MLFF structure (orange), with a connecting line indicating the degree of agreement. The consistently short line segments demonstrate that the MLFF reproduces DFT relaxation behavior with high accuracy. Some representative examples of MLFF-based defect structure optimizations are shown in \textbf{Figure~\ref{fig:gga_m3gnet_model_1}(d--f)} and \textbf{Figure~\ref{fig:hse_m3gnet_model_old_1}(d--f)}  of the SI.
 \\

\begin{figure}[h]
  \centering
  \includegraphics[width=\columnwidth]{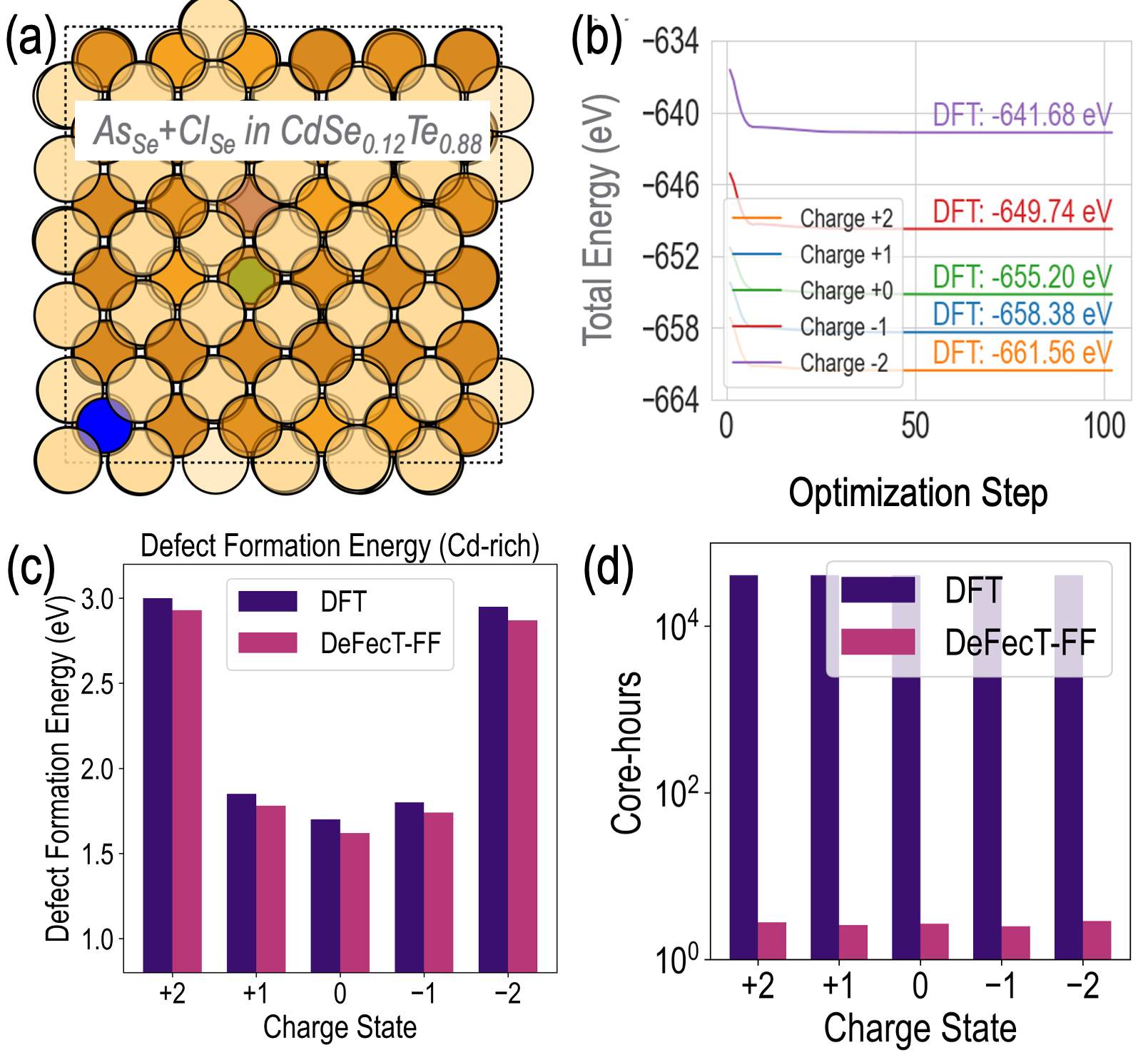}
\caption{Benchmarking \texttt{DeFecT-FF} for a selected defect complex in CdSe$_{0.12}$Te$_{0.88}$.
(a) Visualization of the As$_\mathrm{Se}$ + Cl$_\mathrm{Se}$ complex in the CdSe$_{0.12}$Te$_{0.88}$ alloy supercell. (b) Total energy relaxation profiles for different charge states, comparing the converged DFT energies with the \texttt{DeFecT-FF}–relaxed energies. (c) Defect formation energies under Cd-rich conditions for charge states $+2$ to $-2$, showing close agreement between DFT and \texttt{DeFecT-FF} predictions. (d) Computational cost, measured in core-hours, highlighting the significant reduction in wall-time achieved when using \texttt{DeFecT-FF} instead of full DFT relaxations.}
\label{fig:defect_complex}
\end{figure}

\begin{figure}[h]
  \centering
  \includegraphics[width=\columnwidth]{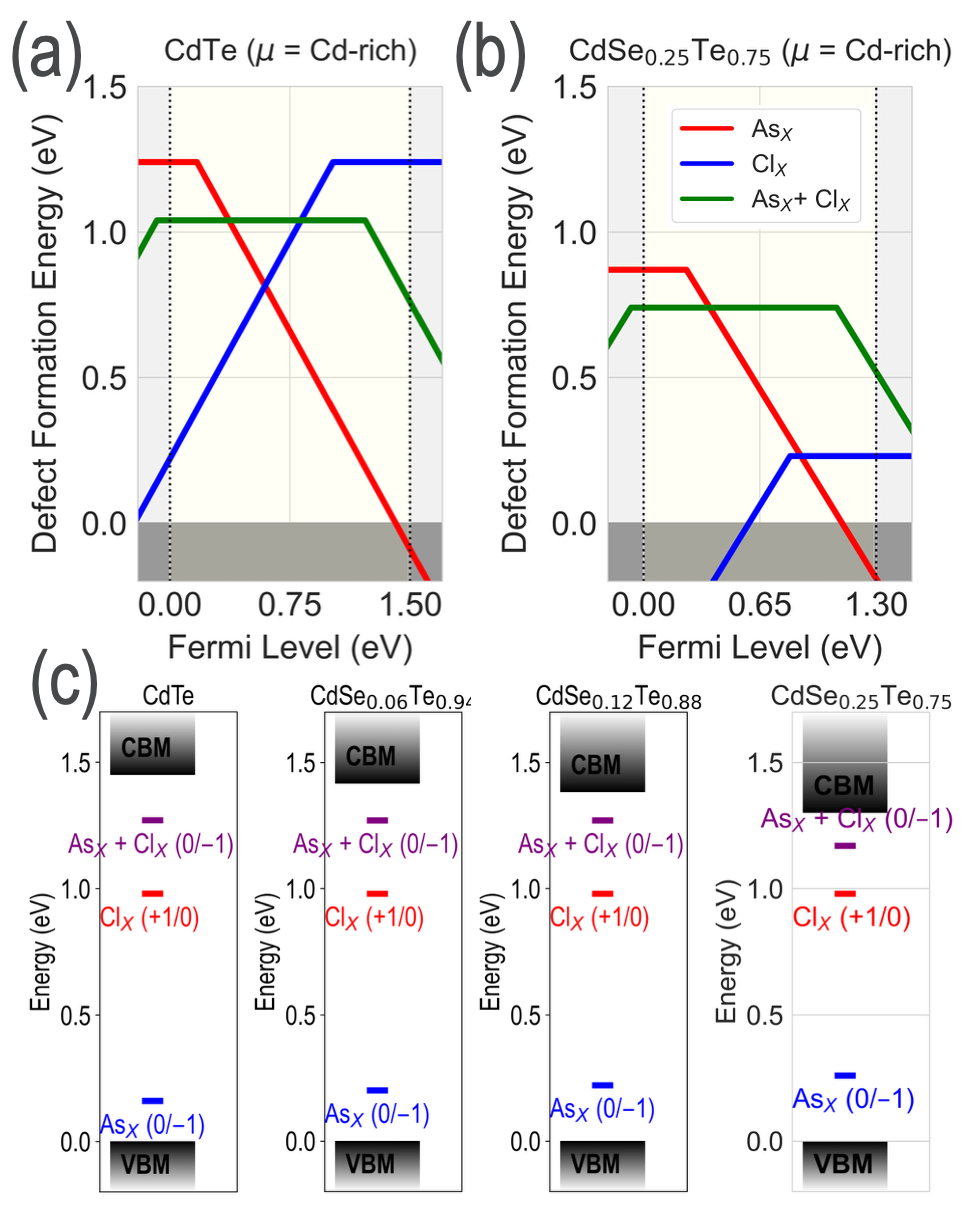}
\caption{Defect formation energy diagrams for As$_{X}$, Cl$_{X}$, and As$_{X}$+Cl$_{X}$ defects in (a) CdTe and (b) CdSe$_{0.25}$Te$_{0.75}$, under Cd-rich conditions; X = Te or Se. (c) Defect charge transition levels for As$_{X}$, Cl$_{X}$, and As$_{X}$+Cl$_{X}$, computed for different CdSe$_x$Te$_{1-x}$ compositions ($x = 0.0$, 0.06, 0.12, 0.25). Blue lines indicate the As$_{X}$ (0/–1) acceptor level, red lines show the Cl$_{X}$ (+1/0) donor level, and purple lines show the As$_{X}$+Cl$_{X}$ (0/–1) level. For each compound, the VBM is placed at E$_F$ = 0 eV and the CBM (conduction band minimum) is placed at the value of the computed band gap. All results are from HSE06+SOC calculations performed after \texttt{DeFecT-FF} optimization.}
\label{fig:cdsete_mlff}
\end{figure}

The MLFF achieves comparable accuracy across the different material families in our chemical space, as summarized in \textbf{Table~\ref{table:rmse_by_composition}}. Binary compounds exhibit slightly lower RMSE (7--8 meV/atom) compared to ternary alloys (8--10 meV/atom), reflecting the additional structural complexity from compositional disorder. Since the charge correction (which depends on the dielectric constant) is computed from DFT and applied separately, variations in dielectric properties across materials do not impact the MLFF's structural prediction accuracy. Ultimately, there are a sufficient number of bulk and defect configurations from different compositions in the training dataset to ensure accurate predictions across the entire space. \\

We note that separate MLFF models are trained for each charge state (from $q = +2$ to $-2$), with every model learning the configuration-to-energy mapping from the DFT geometry relaxation performed at that specific charge state. Electrostatic effects are thus implicitly encoded in the training data. Long-range periodic image interactions are corrected using the Freysoldt scheme~\cite{Freysoldt2014-en} with DFT-derived quantities. The MLFF captures geometry-driven relaxations through local atomic descriptors, while the electronic structure (band edges, charge corrections, SOC effects) is always determined from the final HSE06+SOC single-point calculation. For defects with delocalized charge densities near band edges, the MLFF still provides accurate structural relaxation; however, it should be noted that the electronic characterization of such shallow defects relies entirely on the DFT step rather than the MLFF prediction. MLFF-driven geometry optimization at different charge states significantly reduces the DFT expense, but the subsequent high-fidelity calculation is necessary to obtain the final defect formation energies. \\

The total computational investment for developing DeFecT-FF includes DFT data generation and MLFF training. The HSE06 dataset generation required approximately 20 million core-hours across all compositions and charge states. Training each charge-state-specific M3GNet~\cite{m3gnet} model required approximately 8--12 GPU-hours on a single NVIDIA A100 GPU. Once trained, the MLFF enables rapid predictions at negligible cost ($\sim$ 0.5 core-hours per defect optimization). The upfront investment in data generation and training is amortized over the large number of subsequent predictions: screening hundreds of defect configurations across multiple compositions requires only days of MLFF computation compared to years of equivalent DFT effort. For users wishing to apply DeFecT-FF to new but related chemistries, fine-tuning the pretrained models on a modest set ($\sim$ 50--100 structures) of new DFT calculations is expected to be sufficient, requiring only a few GPU-hours of additional training. \\

\section{New Predictions with the MLFF Models: Case Studies of Important Defects}

The MLFF models can now be used to optimize any given defect configuration in different charge states with near–hybrid-functional accuracy. Following MLFF optimization, final HSE06+SOC single-point calculations must be performed to obtain reliable defect formation energy diagrams. \textbf{Figure~\ref{fig:workflow}} illustrates the overall workflow of the \texttt{DeFecT-FF} framework \cite{defect-ff} in determining the lowest energy symmetry-broken defect configuration followed by a high-fidelity understanding of the defect thermodynamics. \textbf{Figure~\ref{fig:defectff_nanoHUB}} shows the workflow of the \texttt{DeFecT-FF} \cite{defect-ff} web tool we created on the nanoHUB platform to enable efficient creation of defect structures, MLFF optimization, and visualization of defect formation energy diagrams. In the next subsections, we present a few case studies demonstrating the application of this workflow to determine the relative stability and charge transition levels of important defects in selected Cd/Zn-Te/Se/S compositions which were not entirely part of the MLFF training dataset. \\

\subsection{As+Cl Defect Complex in CdSe$_{0.12}$Te$_{0.88}$}

As is a commonly used p-type dopant in Se-alloyed CdTe and Cl is a common impurity arising from CdCl$_{2}$ treatment, which makes it important to investigate defect complexes of As and Cl in representative CdSeTe compositions. We applied the \texttt{DeFecT-FF} framework to simulate the As$_\mathrm{Se}$ + Cl$_\mathrm{Se}$ substitutional defect complex in the compound CdSe$_{0.12}$Te$_{0.88}$, across five possible charge states. An example configuration is illustrated in \textbf{Figure~\ref{fig:defect_complex}(a)}; Se alloying starting from a 216-atom CdTe cubic supercell is first simulated using the special quasirandom structures (SQS) approach \cite{SQS}, following which As and Cl substitution is incorporated in multiple possible locations to eventually yield the preferred combination. For each charge state between $q=+2$ and $q=-2$, ten symmetry-broken initial structures were generated using the ShakeNBreak protocol~\cite{snb_1}, enabling exploration of a diverse set of competing local geometries. Each of these structures was relaxed using \texttt{DeFecT-FF} to identify the lowest-energy configuration prior to high-fidelity DFT refinement. \\

The total energy relaxation profiles for different charge states are shown in \textbf{Figure~\ref{fig:defect_complex}(b)}. The \texttt{DeFecT-FF} structural optimizations converge smoothly and yield geometries that lie very close to those produced by DFT calculations. After \texttt{DeFecT-FF} relaxation, a single-shot HSE06+SOC calculation is performed on the predicted lowest-energy geometry for each charge state to accurately compute defect formation energies. The resulting defect formation energies (at E$_{F}$=0) under Cd-rich conditions from DFT and \texttt{DeFecT-FF} are compared in \textbf{Figure~\ref{fig:defect_complex}(c)}. Across all charge states, the agreement is excellent, with typical deviations well below 0.1--0.2 eV. \\

Thus, the \texttt{DeFecT-FF} geometries provide a sufficiently accurate structural foundation for defect thermodynamics for complexes in an alloyed CdSeTe composition. The computational savings are substantial: a single HSE06 relaxation of a charged defect in a 3$\times$3$\times$3 supercell requires approximately 512 cores multiplied by 8 to 9 hours per configuration, corresponding to a total of nearly 4096 core-hours. In contrast, the \texttt{DeFecT-FF} relaxations require only about $(2/60)\times 16$ core-hours per configuration, which is approximately 0.5 core-hours. The speedup therefore exceeds four orders of magnitude, as presented in \textbf{Figure~\ref{fig:defect_complex}(d)}. We expect these trends to hold for all types of defect complexes in alloyed 3$\times$3$\times$3 supercells and there is confidence in \texttt{DeFecT-FF} reaching close agreement with hybrid DFT at a fraction of the cost. \\

\subsection{As and Cl Defects Across CdSe$_{x}$Te$_{1-x}$ Compositions}

Next, we simulated multiple substitutional defects of As and Cl (including complexes) in $3\times3\times3$ supercells of a series of CdSe$_{x}$Te$_{1-x}$ compositions ($x = 0,\,0.06,\,0.12,\,0.25$).  Using the \texttt{Doped} \cite{doped} package, we introduced defects As$_{Te}$, As$_{Se}$, Cl$_{Te}$, Cl$_{Se}$ and the As$_{X}$+Cl$_{X}$ double defect complexes (where $X$ denotes the preferred anion site, Te or Se). Symmetry‐breaking operations were then applied via the ShakeNBreak protocol, enabling the sampling of a diverse set of competing configurations (\textbf{Figure~\ref{fig:snb_pes}}). Hundreds of structures for these substitutional defects across the CdSe$_{x}$Te$_{1-x}$ compounds were relaxed with the \texttt{DeFecT-FF} models for different charge states until the maximum force fell below $<10^{-2}$ eV/\AA. Finally, single‐shot HSE06+SOC calculations were performed to obtain accurate defect formation energy. \\

The band gaps computed using HSE06+SOC (with a modified mixing parameter of $\alpha$=0.31) for CdTe, CdSe$_{0.06}$Te$_{0.94}$, CdSe$_{0.12}$Te$_{0.88}$, and CdSe$_{0.25}$Te$_{0.75}$ are respectively 1.5 eV, 1.41 eV, 1.38 eV, and 1.30 eV; these values are used to place the E$_{F}$ bounds for the defect formation energy diagrams. The VBM for each composition was obtained from the bulk calculation at the HSE+SOC level using a $2\times2\times2$ $k$-mesh for the corresponding 216-atom $3\times3\times3$ supercell. The charge-dependent defect formation energies additionally yield the charge transition levels as described below: 

\[
\varepsilon(q/q')
= \frac{\Delta E_{f}(D^{q};E_{F}=0) - \Delta E_{f}(D^{q'};E_{F}=0)}{q' - q}
\]

This transition level marks the E$_{F}$ position at which charge states $q$ and $q'$ are in equilibrium.  \textbf{Figure~\ref{fig:cdsete_mlff}} presents the defect formation energy diagrams and relevant transition levels for As$_{X}$, Cl$_{X}$, and As$_{X}$+Cl$_{X}$ defects across the CdSe$_{x}$Te$_{1-x}$ series, with E$_{VBM}$ set to 0 eV; \textit{X} represents either Te or Se.  Incorporation of Se is observed to deepen the As$_{X}$ 0/-1 acceptor level despite the band gap going down from CdTe to CdSe$_{0.25}$Te$_{0.75}$, in agreement with recent experimental studies \cite{Nardone}. The Cl$_{X}$ +1/0 donor level remains deep in the band gap in all cases, around 1 eV from the VBM, while the As$_{X}$+Cl$_{X}$ defect complex, interestingly, creates a 0/-1 acceptor level closer to the conduction band edge which becomes shallower with more Se content due to the lowering of the CBM. The defect energy diagrams in \textbf{Figure~\ref{fig:cdsete_mlff}(a-b)} show the prevalence of the neutral state for the defect complex in the band gap, while As$_{X}$ and Cl$_{X}$ respectively create low energy acceptor and donor defects which pin the equilibrium E$_F$ (obtained by applying charge-neutrality conditions) around the middle of the band gap. \\

The case studies in this section serve as out-of-distribution validation tests \cite{OOD}: the defect configurations examined (As+Cl complex in CdSe$_{0.12}$Te$_{0.88}$) were not included in the MLFF training dataset. To further quantify OOD performance, we evaluated the model on two compositions entirely absent from the training set: CdSe$_{0.12}$Te$_{0.88}$ and CdSe$_{0.06}$Te$_{0.94}$. As shown in \textbf{Table~\ref{tab:ood_rmse}}, the RMSE values (12--13 meV/atom) are moderately higher than in-distribution errors but confirm reasonable generalization. We note that systematic uncertainty quantification for the deployed MLFF models (e.g., via ensemble predictions or Monte Carlo dropout) remains a direction for future development.

\subsection{Native Defects and Nitrogen Impurities in ZnTe}

\begin{figure}[t]
\centering
\includegraphics[width=\columnwidth]{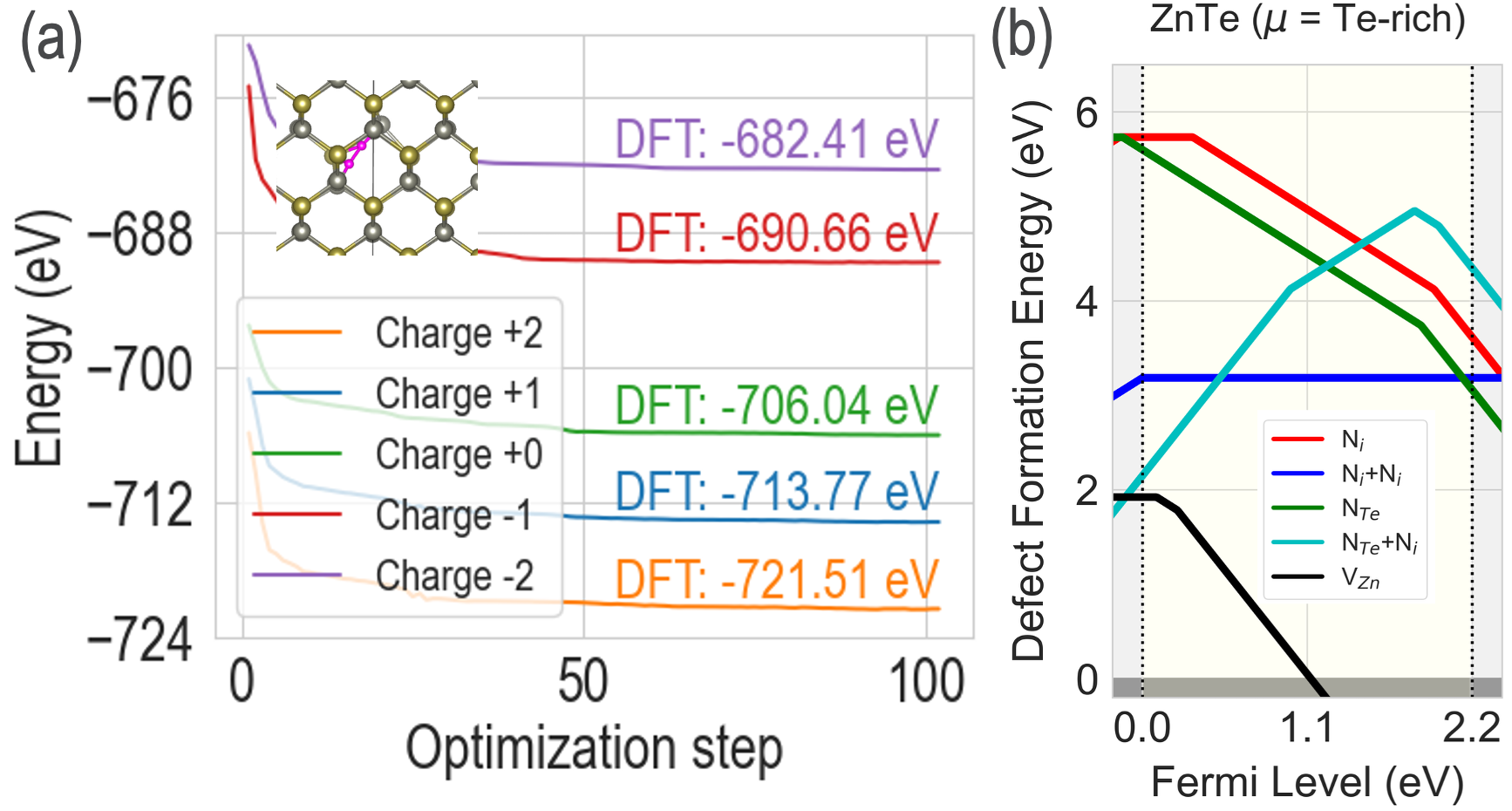}
\caption{ (a) Energy as a function of optimization steps during \texttt{DeFecT-FF} relaxation of a ZnTe supercell with the double N interstitial (N$_i$+N$_i$) defect complex. The inset shows the relaxed configuration with N atoms in red. (b) Defect formation energies in ZnTe under Te-rich conditions computed using HSE06+SOC on top of the HSE-MLFF optimized configurations.}
\label{fig:znte_mlff}
\end{figure}

Motivated by experimental evidence from X-ray photoelectron spectroscopy (XPS) \cite{f__a__stevie_72130f8a, janet_mahoney_e1a9d305, artur_born_2c82645b, gustavo_lanza_7007a95e, hui_chen_dca65eb5, haipeng_xie_92ab6279} indicating N incorporation in ZnTe \cite{ji_ha_lee_5db163c4, lijuan_zhao_6c062f36, yingrui_li_bb64582f, titao_li_7535f555, daniele_dragoni_41c08f2d, eduardo_men_ndez_proupin_dae1679b, ethan_berger_ad4781d2}, we employed the \texttt{DeFecT-FF} workflow to systematically investigate both native point defects and N-related defects in ZnTe. A $3\times3\times3$ ZnTe supercell (cubic zincblende phase) was first fully relaxed using the HSE06 functional prior to defect introduction; its band gap was computed to be 2.2 eV from HSE06+SOC. The defect set included vacancies, interstitials, and antisite defects (V$_{Zn}$, V$_{Te}$, Zn$_{i}$, Te$_{i}$, Zn$_{Te}$, Te$_{Zn}$), as well as the following N defects: N$_{i}$, N$_{Te}$, N$_{i}$+N$_{i}$, and N$_{Te}$+N$_{i}$. To ensure thorough exploration of the potential energy landscape, we applied ShakeNBreak \cite{snb_1, snb_2} to induce perturbations, enabling the sampling of a diverse set of competing configurations. Among the hundreds of N-related configurations evaluated, the N$_{i}$+N$_{i}$ defect complex emerged as the most energetically favorable, a finding further validated through additional HSE06+SOC calculations. \textbf{Figure~\ref{fig:znte_mlff}(a)} illustrates the \texttt{DeFecT-FF} structural optimization and energy convergence for N$_{i}$+N$_{i}$ in ZnTe, and \textbf{Figure~\ref{fig:znte_mlff}(b)} presents the HSE06+SOC computed defect formation energy diagram for different N-related defects in ZnTe. \\

To clarify the role of the MLFF in the defect formation energy workflow: the primary computational bottleneck is the geometry optimization of defect supercells, which requires many ionic relaxation steps using the HSE06 functional (requiring roughly 4,000 core-hours per defect). \texttt{DeFecT-FF} replaces this step with a fast MLFF-based relaxation ($\sim$ 0.5 core-hours), after which a single-point HSE06+SOC calculation is performed on the optimized geometry. The band gap and valence band edge are obtained from DFT on the bulk semiconductor and only need to be computed once per composition. These values, along with available chemical potentials, are combined with MLFF-predicted energies and assumed charge correction energies to obtain MLFF-based defect formation energies. Thus, the MLFF eliminates the need for iterative and expensive DFT defect geometry optimization while enabling rapid screening of many competing configurations, reducing the total cost by over four orders of magnitude. \\

\begin{figure}[t]
\centering
\includegraphics[width=\columnwidth]{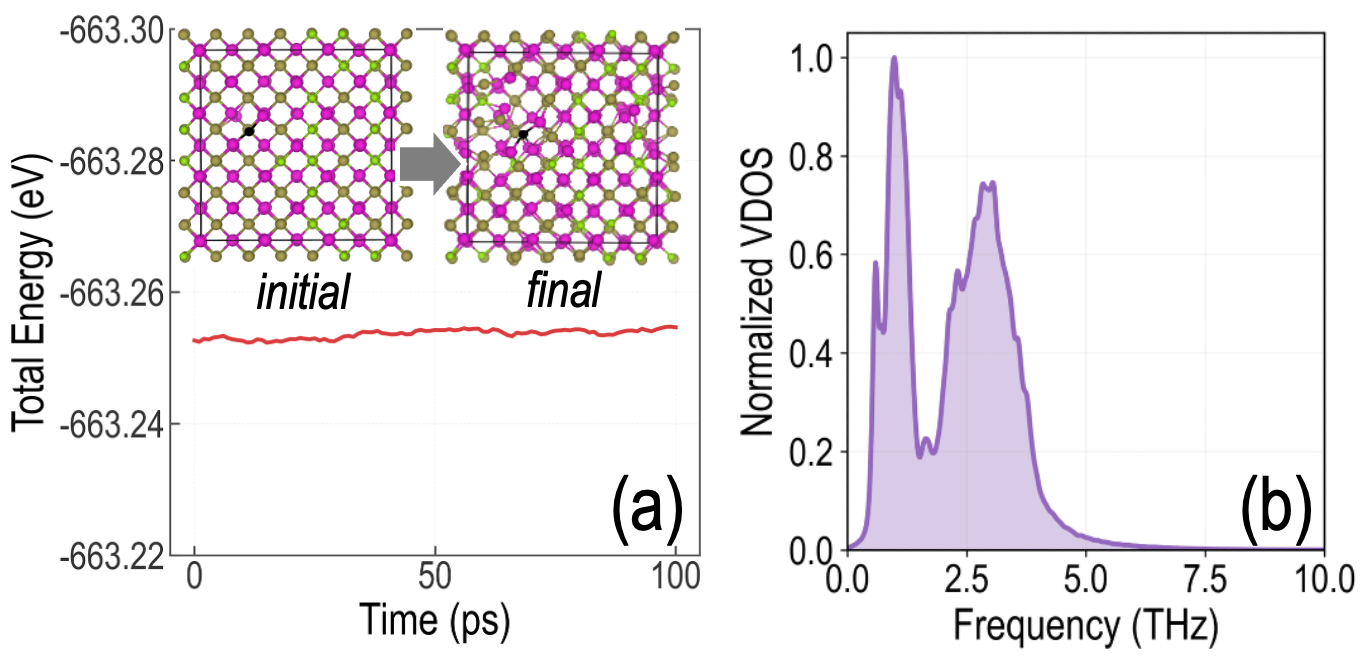}
\caption{(a) Total energy as a function of simulation time for a microcanonical (NVE) molecular dynamics trajectory of the As$_{Te}$ defect in CdTe, demonstrating excellent energy conservation over a 100~ps run. The inset shows the atomic structure before (\textit{initial}) and after (\textit{final}) MD simulation, with the defect site highlighted. (b) Normalized vibrational density of states (VDOS) obtained from the velocity autocorrelation function of the same trajectory, revealing dominant low-frequency vibrational modes below 4~THz that are characteristic of defect-localized and heavy-atom vibrations.}
\label{fig:md_simulations}
\end{figure}

\subsection{Simulating Defects at Finite Temperature}

Defect formation energies in semiconductors are typically evaluated at $T = 0$~K using DFT, neglecting vibrational entropy effects that can become important under realistic growth and operating conditions. While a full finite-temperature free-energy treatment is beyond the scope of this work, we demonstrate that \texttt{DeFecT-FF} enables finite-temperature molecular dynamics (MD) \cite{MD_1, MD_2} simulations that provide a physically grounded pathway toward incorporating such effects in future studies. As a representative example, we consider the As$_{Te}$ defect in CdTe. The defect structure was first relaxed using \texttt{DeFecT-FF}, followed by finite-temperature microcanonical (NVE) ensemble MD simulations for the neutral charge state ($q=0$) at $T=300$~K using a 1~fs time step and a total simulation length of 100{,}000 steps \cite{MD_3, MD_4, MD_5}. \textbf{Figure~\ref{fig:md_simulations}(a)} shows the energy as a function of simulation time, demonstrating excellent energy conservation over the 100~ps trajectory and confirming the numerical stability of \texttt{DeFecT-FF}. \\

Beyond validating stability, finite-temperature MD offers key physical advantages over static relaxations by enabling symmetry-breaking driven by thermal fluctuations and realistic atomic motion.During the MD simulations, atomic velocities $v_i(t)$ were recorded at each time step, where $i$ labels atoms in the supercell and $N$ is the total number of atoms. From these trajectories, we computed the velocity autocorrelation function as follows \cite{Rahman_Sun}:
\begin{equation}
C_{vv}(t) =
\frac{1}{3N}
\sum_{i=1}^{N}
\langle
v_i(0) \cdot v_i(t)
\rangle,
\end{equation}
and obtained the vibrational density of states (VDOS), $g(\omega)$, using Fourier transformation:
\begin{equation}
g(\omega) =
\int_{0}^{\infty}
C_{vv}(t)\,
e^{-i \omega t}\,
dt,
\end{equation}
Here, $\omega = 2\pi f$ is the vibrational angular frequency. \textbf{Figure~\ref{fig:md_simulations}(b)} shows the normalized VDOS derived from the MD trajectory. The dominant low-frequency modes below approximately 4~THz are characteristic of defect-localized vibrations and reflect vibrational softening induced by the As$_{Te}$ defect and the presence of heavy atoms in CdTe. Beyond confirming numerical stability, the finite-temperature MD simulation provides a few physical insights and opportunities: (i) the VDOS reveals defect-specific vibrational signatures, including low-frequency modes characteristic of defect-localized vibrations; (ii) thermal fluctuations enable dynamical exploration of the local configurational landscape \cite{Freysoldt2014-en}, potentially accessing lower-energy structures through symmetry-breaking pathways inaccessible at 0 K; and (iii) the VDOS provides direct access to vibrational entropy contributions needed for computing temperature-dependent defect formation free energies, establishing a foundation for future investigations of defect thermodynamics under realistic conditions which will be addressed in follow-up contributions from our group. \\

\section{Conclusions}

CdSe$_x$Te$_{1-x}$ solar cells are fundamentally constrained by defect physics: deep-level nonradiative centers from native defects and impurities limit open-circuit voltage, dopants such as Cu and As often lead to unhelpful complexes, and extended defects at interfaces and grain boundaries act as sinks for charge and sites for defect clustering. While hybrid-functional DFT remains the gold standard for resolving these mechanisms, its cost prevents exhaustive exploration across alloy compositions, charge states, and structural motifs. To overcome these barriers, we introduced the \texttt{DeFecT-FF} framework in this work, a crystal graph-based active learning-driven MLFF model trained on data from both semi-local GGA and hybrid HSE06 calculations for thousands of charged and neutral structures spanning the Cd/Zn–S/Se/Te chemical space, with a wide variety of native and extrinsic defects and defect complexes considered. \texttt{DeFecT-FF} predicts energies and forces across charge states, enabling rapid geometry optimization and defect formation energy evaluation. We demonstrated the utility of these models by identifying low energy configurations of device-relevant defects and performing HSE06+SOC calculations to understand their energetics and defect levels. \\

In practice, the \texttt{DeFecT-FF} framework reduces single defect optimization time from at least $\sim$8--9\,h (HSE06) to $\sim$1--2\,min while retaining near-DFT accuracy, transforming comprehensive, composition- and charge-resolved defect surveys from intractable to routine. The term ``near-DFT accuracy'' refers specifically to the MLFF achieving RMSE values of 5--10 meV/atom in crystal formation energy and $<$0.20 eV in defect formation energy relative to HSE06 DFT. ``High-fidelity'' refers to the use of the HSE06+SOC for final single-point calculations. \\

We have deployed this framework as part of a Jupyter notebook-based nanoHUB tool which will allow users to upload CIF files of Cd/Zn-Te/Se/S structures, auto-generate relevant defects or complexes, and compute their defect formation energies as functions of Fermi level and chemical potentials conditions, bypassing expensive first principles workflows. Together, these advances provide a scalable, charge-aware pathway to map defect landscapes in chemistries relevant to CdSeTe solar cell devices and beyond, accelerating the dopant/process optimization and ultimately closing the voltage deficit in this important thin-film photovoltaic platform. \\

\section*{Conflicts of Interest}
\vspace{0.2cm}
There are no conflicts to declare. \\

\section*{Data Availability}
All raw and processed data supporting this work, namely atomic structures, total energies, forces, and stresses, and the trained MLFF models used in this study, are publicly accessible via a nanoHUB web application at \url{https://nanohub.org/tools/defectdatabase}. This tool enables community reuse and verification through a point-and-click interface. Within the same tool, users can:
\begin{itemize}
  \item \textbf{Provide inputs:} Upload a crystallographic file (e.g., CIF or POSCAR), select native or extrinsic defect types (vacancy, interstitial, substitutional, or complexes).
  \item \textbf{Run computations:} Perform MLFF-based geometry relaxation in minutes and obtain optimized energies to identify lowest energy configurations. 
  \item \textbf{Retrieve outputs:} Download relaxed structures, per-defect energies, and tabulated CSV summaries, along with run logs for reproducibility.
  \item \textbf{Intended scope:} Rapid screening and hypothesis testing; device-critical cases should be validated with targeted HSE06(+SOC) calculations. \\
\end{itemize}

\section*{Funding Statement}

This material is based upon work supported by the U.S. Department of Energy’s Office of Energy Efficiency and Renewable Energy (EERE) under the Solar Energy Technology Office (SETO) Award Number DE-0009332. Funding for this work was also provided by the Alliance for Sustainable Energy, LLC, Managing and Operating Contractor for the National Renewable Energy Laboratory for the U.S. DOE, and was supported in part by EERE under SETO Award Number 37989. A.M.K. additionally acknowledges support from Argonne National Laboratory under sub-contracts 21090590 and 22057223, from DOE EERE. This research used resources from the the Center for Nanoscale Materials (CNM) at Argonne National Laboratory. Work performed at the CNM, a U.S. Department of Energy Office of Science User Facility, was supported by the U.S. DOE, Office of Basic Energy Sciences, under Contract No. DE-AC02-06CH11357. This work also utilized the Anvil cluster at Purdue through allocation MAT230030 from the Advanced Cyberinfrastructure Coordination Ecosystem: Services \& Support (ACCESS) program, which is supported by U.S. National Science Foundation grants 2138259, 2138286, 2138307, 2137603, and 2138296. \\

\section*{Acknowledgments}

The authors would like to acknowledge discussions with Dr. Mariana Bertoni at Arizona State University, Dr. Yanfa Yan at University of Toledo, Dr. Mike Scarpulla at University of Utah, and researchers at the National Renewable Energy Laboratory. We also acknowledge the Rosen Center for Advanced Computing (RCAC) clusters at Purdue University for further computational support. \\

\section*{Author Contributions Statement}

A.M.-K. conceived and planned the research project and procured research funding. DFT computations and MLFF training tasks were performed by M.H.R. and M.B. M.H.R. took the lead on writing; M.B. and A.M.-K. contributed in editing and shaping the manuscript. \\

\bibliographystyle{rsc}
\bibliography{reference}

\pagenumbering{gobble}
\thispagestyle{empty}

\onecolumn
\setcounter{figure}{0}
\setcounter{table}{0}
\renewcommand{\thetable}{S\Roman{table}}
\renewcommand\thefigure{S\arabic{figure}}
\begin{center}
\vspace{0.5cm}
\Large
\textbf{Supplemental material to "DeFecT-FF: A Machine Learning Force Field Framework for High Throughput Defect Modeling in CdTe-Based Solar Cells"}
\vspace{0.5cm}
\large \\
Md Habibur Rahman,\textsuperscript{1)} Maitreyo Biswas,\textsuperscript{1)} and Arun Mannodi-Kanakkithodi \textsuperscript{1, a)}
\vspace{0.1cm}
\normalsize
\\
\textsuperscript{1}\textit{School of Materials Engineering, Purdue University, West Lafayette, Indiana 47907, USA}\\
\end{center}
\vspace{5em} 
\footnotetext{\textsuperscript{a}amannodi@purdue.edu\hspace{0.3cm}}
\begin{table}[h]
    \centering
    \caption{
    Enumeration of defects across binary and alloyed Cd/Zn–Te/Se compositions. Binary compounds (CdTe, CdSe, ZnTe) contain two atomic species and therefore host 6 native defect types. Alloyed compounds contain three distinct ionic species, yielding 12 native defect types. Extrinsic defect species (Cu; As, P, N, Sb, Bi; Cl, O) contribute three defect types each (interstitial, cation substitution, anion substitution). For mixed compositions, we enforce a minimum of 10 non-equivalent positions per defect type, even through thousands of distinct symmetry-inequivalent positions exist (e.g., As$_i$ in Cd–Se–Te). Each defect configuration is additionally perturbed using ShakeNBreak \cite{snb_2} to generate at least 15 symmetry-broken initial structures.}
    \label{tab:defect_counts_v1}

    \renewcommand{\arraystretch}{1.15}
    \setlength{\tabcolsep}{6pt}

    \begin{tabular}{lccccccc}
        \hline
        Compound &
        Native &
        Extrinsic &
        Native + Extrinsic &
        Double complexes &
        Total unique defects &
        \makecell{Non-eq\\sites\\($\times$10)} &
        \makecell{Perturbed\\structures\\($\times$15)} \\
        \hline
        CdTe                         & 6  & 24 & 30 & 435 & 465 & - & 6{,}975 \\
        CdSe                         & 6  & 24 & 30 & 435 & 465 & - & 6{,}975 \\
        ZnTe                         & 6  & 24 & 30 & 435 & 465 & - & 6{,}975 \\
        CdSe$_{0.25}$Te$_{0.75}$     & 12 & 32 & 44 & 946 & 990 & 9{,}900 & 148{,}500 \\
        CdSe$_{0.50}$Te$_{0.50}$     & 12 & 32 & 44 & 946 & 990 & 9{,}900 & 148{,}500 \\
        CdSe$_{0.75}$Te$_{0.25}$     & 12 & 32 & 44 & 946 & 990 & 9{,}900 & 148{,}500 \\
        Cd$_{0.25}$Zn$_{0.75}$Te     & 12 & 32 & 44 & 946 & 990 & 9{,}900 & 148{,}500 \\
        Cd$_{0.50}$Zn$_{0.50}$Te     & 12 & 32 & 44 & 946 & 990 & 9{,}900 & 148{,}500 \\
        Cd$_{0.75}$Zn$_{0.25}$Te     & 12 & 32 & 44 & 946 & 990 & 9{,}900 & 148{,}500 \\
        \hline
    \end{tabular}
\end{table}
\begin{table}[h!]
\centering
\renewcommand{\arraystretch}{1.2}
\begin{tabular}{|c|c|c|}
\hline
\textbf{Composition} & \textbf{PBE (\AA)} & \textbf{HSE06 (\AA)} \\
\hline
\multicolumn{3}{|c|}{\textbf{CdSe$_x$Te$_{1-x}$}} \\
\hline
CdTe ($x=0.00$)               & 19.88 & 19.72 \\
CdSe$_{0.25}$Te$_{0.75}$      & 19.57 & 19.34 \\
CdSe$_{0.50}$Te$_{0.50}$      & 19.25 & 19.02 \\
CdSe$_{0.75}$Te$_{0.25}$      & 18.94 & 18.71 \\
CdSe ($x=1.00$)               & 18.63 & 18.45 \\
\hline
\multicolumn{3}{|c|}{\textbf{Cd$_x$Zn$_{1-x}$Te}} \\
\hline
CdTe ($x=1.00$)               & 19.88 & 19.72 \\
Cd$_{0.75}$Zn$_{0.25}$Te      & 19.56 & 19.35 \\
Cd$_{0.50}$Zn$_{0.50}$Te      & 19.21 & 19.01 \\
Cd$_{0.25}$Zn$_{0.75}$Te      & 18.85 & 18.72 \\
ZnTe ($x=0.00$)               & 18.52 & 18.42 \\
\hline
\end{tabular}
\caption{Optimized supercell lattice parameters (in \AA) for CdSe$_x$Te$_{1-x}$ and Cd$_x$Zn$_{1-x}$Te compounds computed from PBE and HSE06.}
\label{tab:lattice_parameters}
\end{table}
\begin{figure*}[t]
\centering
\includegraphics[width=.9\linewidth]{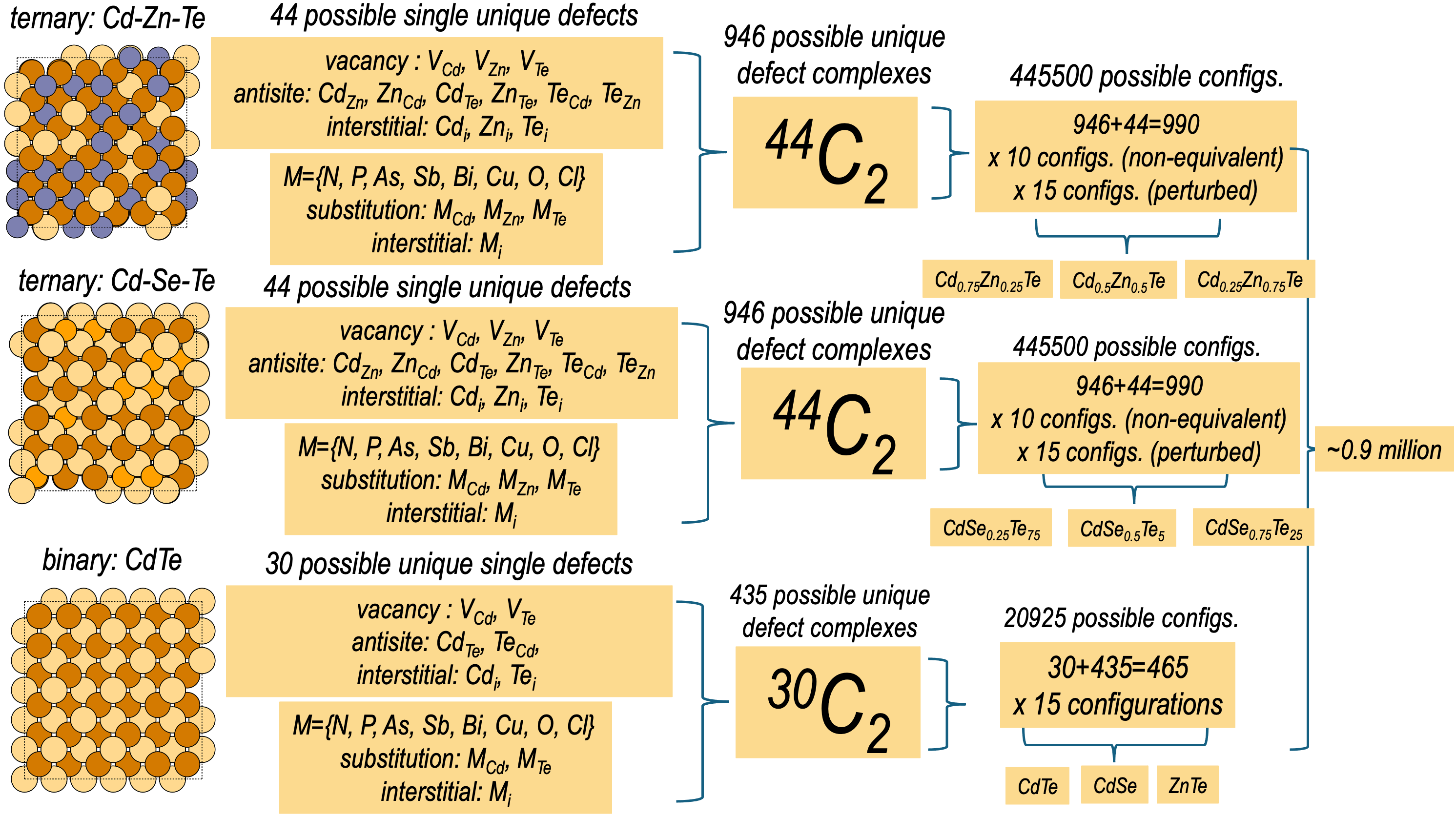}
\caption{\label{fig:chemical_sapce} Schematic illustration of the combinatorial expansion of the full Cd/Zn--Te/Se defect chemical space across binary and ternary compositions. Each ternary alloy (Cd--Zn--Te and Cd--Se--Te) hosts 44 unique single-defect configurations arising from vacancies, antisites, interstitials, and eight extrinsic defects (N, P, As, Sb, Bi, Cu, O, Cl). These yield ${}^{44}C_{2}=946$ unique defect complexes. Combining all single defects and complexes gives 990 distinct defect types per composition, which are further expanded by sampling at least 10 non-equivalent lattice positions per defect and generating 15 ShakeNBreak \cite{snb_2} symmetry-broken perturbations per configuration, resulting in approximately $\sim 1.495\times10^{5}$ total structures per ternary alloy. For the binary compound CdTe (and similarly CdSe and ZnTe), 30 unique single defects generate ${}^{30}C_{2}=435$ complexes, yielding unique 465 defects, each further expanded through 15 perturbed initial structures to give 6{,}975 configurations.}
\end{figure*}

\section*{Detailed Description of the DFT Datasets}
\label{sec:initial_dft_datasets}

The full DFT dataset used in this work includes two level of theories: a broad GGA--PBE dataset containing tens of thousands of bulk and defect configurations, and subset of PBE data refinement using the HSE06 hybrid functional. The majority of the PBE dataset was collected from previously published works from our group covering defects in a wide range of II--VI semiconductors \cite{Mannodi-ZB, Mannodi_Toriyama_2020, Mannodi-Kanakkithodi2022-ck, Rahman_Gollapalli}. These structures were combined with a significant number of new PBE calculations, particularly for alloy compositions and defect configurations that were not part of prior studies. Our chemical space spans binary Cd/Zn--Te/Se/S compounds, selected ternaries, and several quaternary alloys. \textbf{Table~\ref{table:bulk_compounds_list}} lists the full set of bulk compounds included in this study. Each composition is simulated using both $2\times 2\times 2$ (64-atom) zincblende supercells generated using the special quasirandom structure (SQS) approach~\cite{SQS}. For each composition, the mixing fraction $x$ is systematically varied in increments of $0.125$ (i.e., $n/8$ for $n\in\{0,\dots,8\}$), yielding a total of 81 unique compositions~\cite{Mannodi-ZB}. \\

\begin{table}[h]
    \centering
    \caption{
    Summary of bulk compounds included in the PBE/HSE dataset. The full set spans binary, ternary, and quaternary compositions in the Cd/Zn--Te/Se/S chemical space.
    }
    \label{table:bulk_compounds_list}
    \begin{tabular}{ll}
    \hline
    Category & Compounds \\
    \hline
    Binary & CdTe, CdSe, CdS, ZnTe, ZnSe, ZnS \\
    Ternary alloys &
    CdSe$_x$Te$_{1-x}$, CdS$_x$Se$_{1-x}$,
    ZnSe$_x$Te$_{1-x}$, ZnS$_x$Se$_{1-x}$ \\
    Quaternary alloys &
    Cd$_x$Zn$_{1-x}$S, Cd$_x$Zn$_{1-x}$Se,
    Cd$_x$Zn$_{1-x}$Te,
    Cd$_{0.5}$Zn$_{0.5}$S$_x$Se$_{1-x}$,
    Cd$_{0.5}$Zn$_{0.5}$Se$_x$Te$_{1-x}$ \\
    \hline
    \end{tabular}
\end{table}

Across all compositions and supercell sizes, the PBE dataset contains more than 10,000 bulk
structures, including not only the fully relaxed configurations but also all intermediate
snapshots along the geometry-optimization trajectories. Each structure is labeled by its crystal formation energy (CFE), defined for a general composition Cd$_a$Se$_b$Te$_c$ as:

\begin{equation}
    CFE = 
    \frac{E(\text{Cd}_a\text{Se}_b\text{Te}_c)
    - aE(\text{Cd}) - bE(\text{Se}) - cE(\text{Te})}
    {N_{\text{atoms}}}.
\end{equation}

Here, $E(\text{Cd$_{a}$Se$_{b}$Te$_{c}$})$ is the total DFT energy of the supercell containing \textit{a} atoms of Cd, \textit{b} atoms of Se, and \textit{c} atoms of Te that is necessary to simulate the CdSe$_{x}$Te$_{1-x}$ composition. $E(\text{Cd})$, $E(\text{Se})$, and $E(\text{Te})$ are respectively the per-atom energies of Cd, Se, and Te in their known elemental standard states, and $N_{\text{atoms}}$ = \textit{a+b+c} is the total number of atoms in the supercell. 

\begin{figure}[!htbp]
  \centering
  \includegraphics[width=\textwidth]{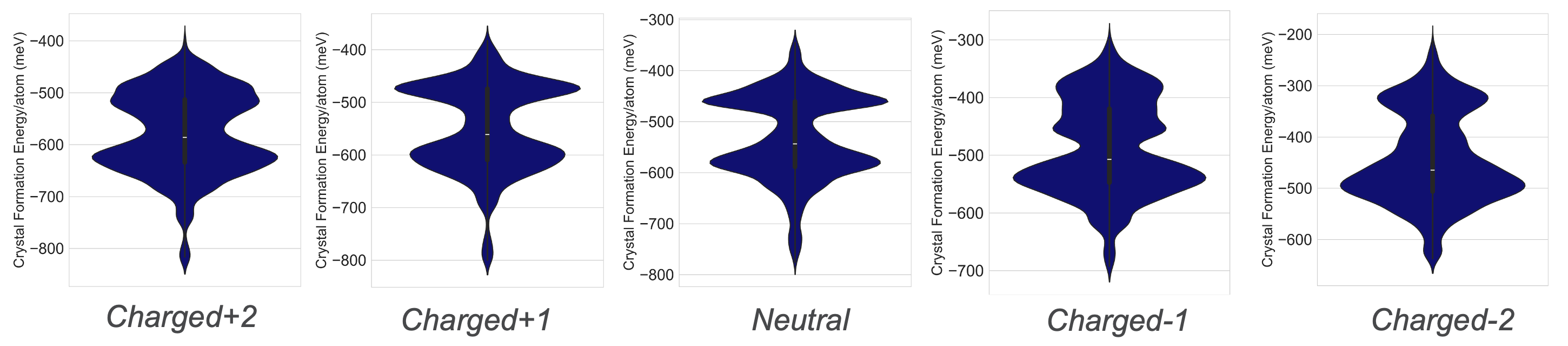}
  \captionof{figure}{Violin distributions of crystal formation energy per atom (meV) for charge states $+2$, $+1$, $0$ (neutral), $-1$, and $-2$ in the GGA-PBE dataset.}
  \label{fig:pbe-violins}
\end{figure}

\begin{figure*}[t]
\centering
\includegraphics[width=.9\linewidth]{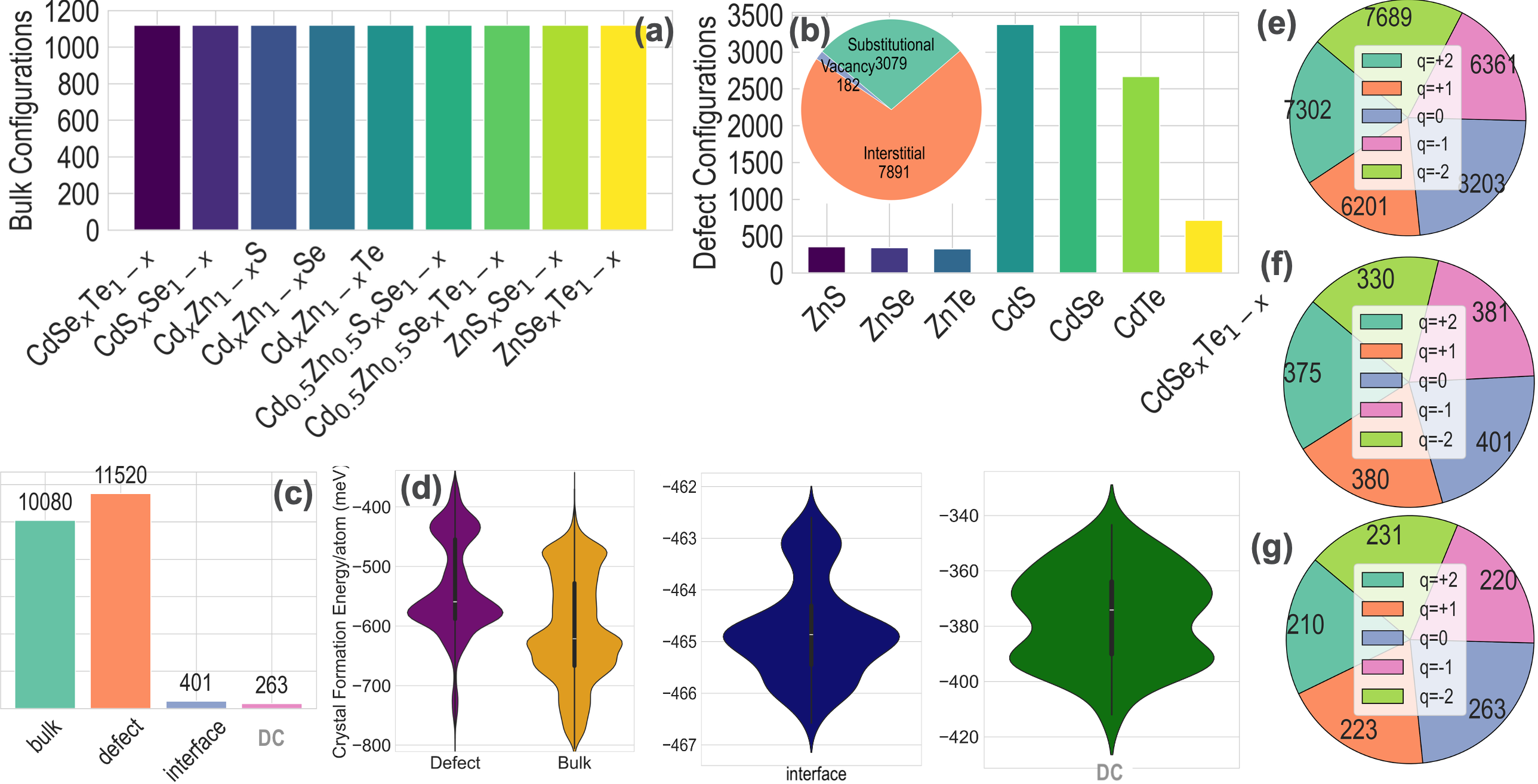}
\caption{\label{fig:initial_pbe_dataset} Statistics of the GGA-PBE dataset: (a) Number of bulk configurations corresponding to CdSe$_{x}$Te$_{1-x}$, CdS$_{x}$Se$_{1-x}$, Cd$_{x}$Zn$_{1-x}$S, Cd$_{x}$Zn$_{1-x}$Se, Cd$_{x}$Zn$_{1-x}$Te, Cd$_{0.5}$Zn$_{0.5}$S$_{x}$Se$_{1-x}$, Cd$_{0.5}$Zn$_{0.5}$Se$_{x}$Te$_{1-x}$, ZnS$_{x}$Se$_{1-x}$, and ZnSe$_{x}$Te$_{1-x}$ compositions. (b) Number of neutral defect configurations in CdS, CdSe, CdTe, ZnS, ZnSe, ZnTe, and different CdSe$_{x}$Te$_{1-x}$ compositions, with the inset showing the distribution of vacancy, substitutional, and interstitial defects across the dataset. (c) Bar chart showing the total number of bulk, defect, interface, and dislocation core (DC) configurations in the dataset. (d) Violin plots showing the distribution of crystal formation energy across all the defect, bulk, interface and DC structures. Inside each violin, a mini box plot shows the median (central line), quartile, and range (whiskers) (e, f, g) Distribution of defect configurations for different charge states (q = +2, +1, 0, -1, and -2) in bulk defect, interface, and DC configurations, respectively.  }
\end{figure*}

The defect dataset contains native vacancies, interstitials, antisites, extrinsic interstitials and substitutional dopants (Cu, As, P, N, Sb, Cl, O), and selected complexes across multiple compositions. All defect calculations were performed in either $2\times 2\times 2$ or $3\times 3\times 3$ supercells, and multiple charge states were considered. Additionally, we incorporated CdTe dislocation–core structural models derived from prior STEM-based work \cite{Fatih-CdTe-GB}. These structures correspond to a 4.8$^\circ$ low-angle (110)$\parallel$(110) tilt boundary, in which the extended defect is represented by a periodic array of Lomer-type dislocation cores. Two experimentally observed core reconstructions were simulated: a \textit{Type-I} core, where the absence of a central Cd column produces a Te-terminated dislocation core with broken Te–Cd bonds that generate deep mid-gap states; and a \textit{Type-II} core, where a restored central Cd column leads to a distinct mixed Cd–Te coordination environment and an electronic state. Together, these two structures capture the essential bonding rearrangements, dangling-bond motifs, and charge perturbations intrinsic to realistic CdTe dislocation cores identified through atomic-resolution STEM imaging~\cite{Fatih-CdTe-GB}. We also constructed a CdTe–ZnTe heterointerface by merging fully relaxed 3$\times$3$\times$3 supercells of CdTe and ZnTe. Small in-plane strains were applied to lattice-match the two materials, followed by volume-conserving structural relaxation to obtain a physically consistent interface. For all new defect calculations in the 3$\times$3$\times$3 binary, alloyed, and interface the \texttt{Doped} package \cite{doped} was used to generate symmetry-broken initial configurations via bond distortions and atomic rattling. In the unified dataset, the label \texttt{bulk} denotes pristine supercells; \texttt{defect} refers to supercells containing vacancies, interstitials, antisites, dopants, or multi-defect complexes; \texttt{interface} corresponds to CdTe–ZnTe heterostructures containing defects at or near the interfacial region; and \texttt{dislocation\_core} designates structures based on the Type-I and Type-II STEM-derived dislocation-core models, with defects introduced within their reconstructed core environments. The complete PBE dataset distributions used for ALIGNN training are reported in \textbf{Figure~\ref{fig:pbe-violins}}, \textbf{Figure~\ref{fig:initial_pbe_dataset}}, and \textbf{Table~\ref{table:initial_pbe_dataset}.} \\

\begin{figure}[!htbp]
  \centering
  \includegraphics[width=\textwidth]{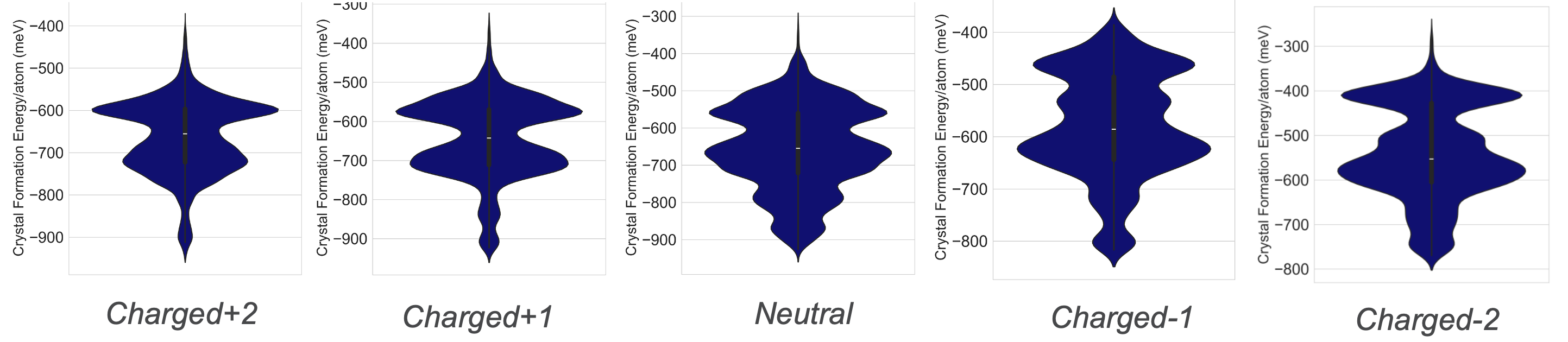}
  \captionof{figure}{Violin distributions of crystal formation energy per atom (meV) for charge states $+2$, $+1$, $0$ (neutral), $-1$, and $-2$ in the HSE06 dataset.}
  \label{fig:hse-violins}
\end{figure}

A curated subset of bulk and defect structures from the PBE dataset was re-optimized using the HSE06 hybrid functional with a mixing parameter of $\alpha = 0.25$. Due to high computational costs, HSE06 calculations were performed only for representative configurations spanning different compositions, and chemical environments. For defect calculations, lattice parameters were updated to the HSE06-optimized bulk volume before relaxation. All hybrid-functional calculations used $\Gamma$-point sampling and reduced plane-wave cutoffs appropriate for large supercells. The distributions of the complete HSE dataset are presented in \textbf{Figure~\ref{fig:hse-violins}} and \textbf{Table~\ref{table:hse_dataset}}. This dataset forms the basis for the MLFF~\cite{ALIGNN} models used in this work.

\begin{table*}[t]
    \centering
    \begin{adjustbox}{width=1\textwidth}
    \begin{tabular}{|c|c|c|}
    \hline
        \textbf{Dataset} & \textbf{Supercell Size} & \textbf{Data Points} \\ \hline
        Bulk dataset from Cd/Zn-Te/Se/S binary, ternary and quaternary alloys & 2$\times$2$\times$2 & 10080   \\ \hline
        Bulk dataset from CdSe$_{x}$Te$_{1-x}$ alloys & 3$\times$3$\times$3 & 26   \\ \hline
        Defect dataset from 6 Cd/Zn-Te/Se/S binary compounds & 2$\times$2$\times$2 & 7302 (q=+2), 6201 (q=+1), 8203 (q=0), 6361 (q=-1), 7689 (q=-2) \\ \hline
        Defect dataset from CdSe$_{x}$Te$_{1-x}$ and Cd$_{x}$Zn$_{1-x}$Te alloys & 3$\times$3$\times$3 & 594 (q=+2), 533 (q=+1), 643 (q=0), 514 (q=-1), 560 (q=-2) \\ \hline
        Defect dataset from CdTe/ZnTe interface & 3$\times$3$\times$6 & 375 (q=+2), 380 (q=+1), 401 (q=0), 381 (q=-1), 330 (q=-2) \\ \hline
        Defect dataset from CdTe dislocation core & -- & 210 (q=+2), 223 (q=+1), 263 (q=0), 220 (q=-1), 231 (q=-2) \\ \hline
    \end{tabular}
    \end{adjustbox}
    \caption{\label{table:initial_pbe_dataset} Number of data points (or structures) in the GGA-PBE dataset corresponding to different types of bulk or defect configurations, supercell sizes, and charge states.}
\end{table*}

\begin{table*}[t]
    \centering
    \begin{adjustbox}{width=1\textwidth}
    \begin{tabular}{|c|c|c|}
    \hline
        \textbf{Dataset} & \textbf{Supercell Size} & \textbf{Data Points} \\ \hline
        Bulk dataset from Cd/Zn-Te/Se/S binary compounds and alloys & 2$\times$2$\times$2 & 5400   \\ \hline
        Defect dataset from 6 binary compounds & 2$\times$2$\times$2 & 4302 (q=+2), 4201 (q=+1), 6203 (q=0), 4361 (q=-1), 4689 (q=-2) \\ \hline
        Defect dataset from CdSe$_{x}$Te$_{1-x}$ and Cd$_{x}$Zn$_{1-x}$Te alloys & 3$\times$3$\times$3 & 371 (q=+2), 333 (q=+1), 402 (q=0), 321 (q=-1), 350 (q=-2) \\ \hline

    \end{tabular}
    \end{adjustbox}
    \caption{\label{table:hse_dataset} Number of data points (or structures) in the HSE06 dataset corresponding to different types of bulk or defect configurations, supercell sizes, and charge states.}
\end{table*}

\begin{figure*}[t]
\centering
\includegraphics[width=.9\linewidth]{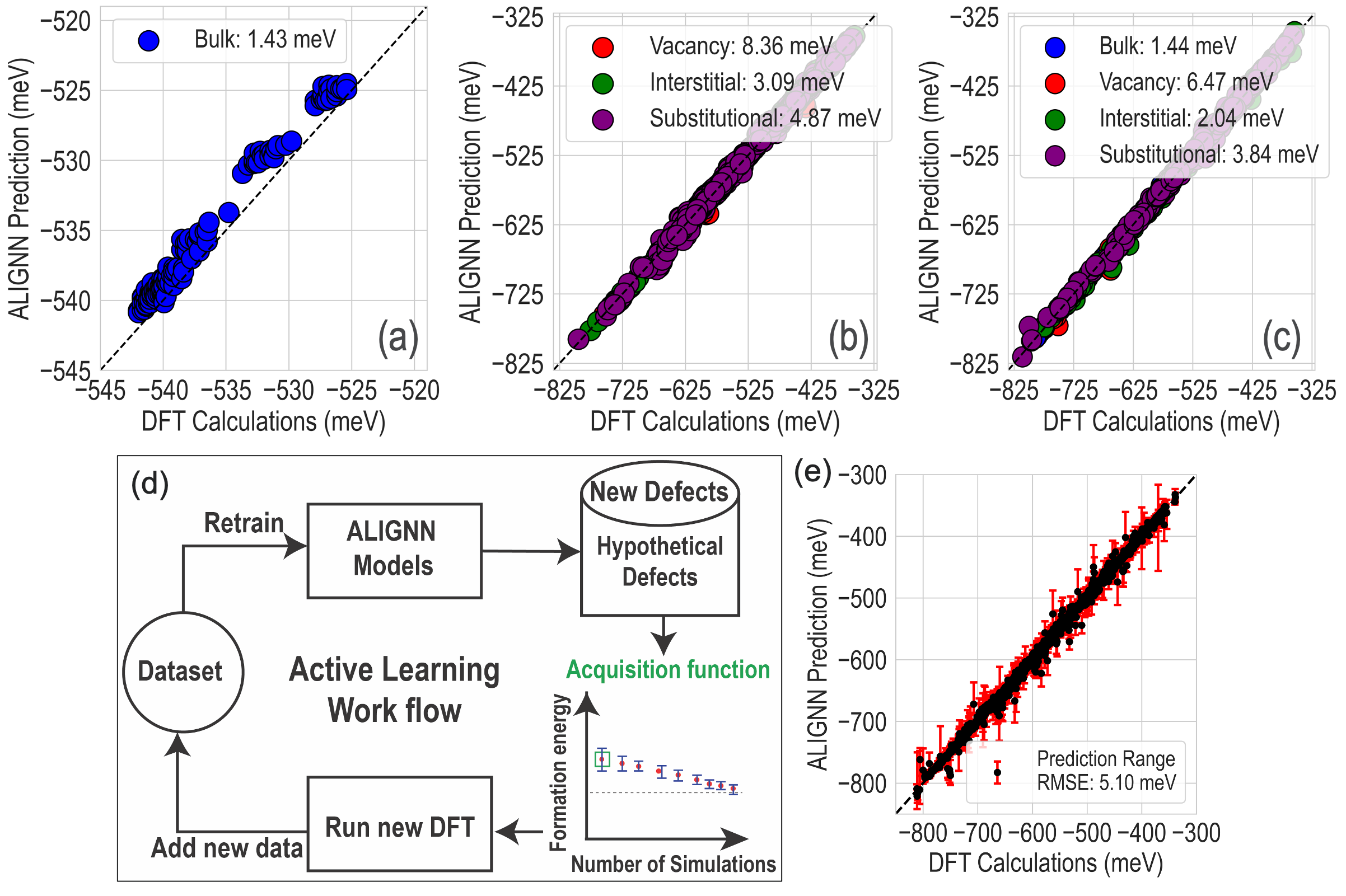}
\caption{\label{fig:al_workflow} Parity plots for ALIGNN models trained on the GGA dataset using (a) only bulk structures, (b) only defect structures, with predictions distinguished in terms of type of defect, and (c) both bulk and defect structures. (d) Active learning (AL) workflow implemented in this work: standard deviation of ALIGNN-predicted formation energy of novel defects is used to determine acquisition functions and identify the next set of DFT simulations to run. The ALIGNN model is then retrained with the new data and predictions are made for the remaining set of unexplored defects. (e) An ALIGNN vs DFT parity plot showing standard deviation in test set prediction across 100 separate models; these error bars are used to calculate the acquisition functions in the AL workflow.}
\end{figure*}

\begin{figure}[!htbp]
    \centering
    \includegraphics[width=.9\linewidth]{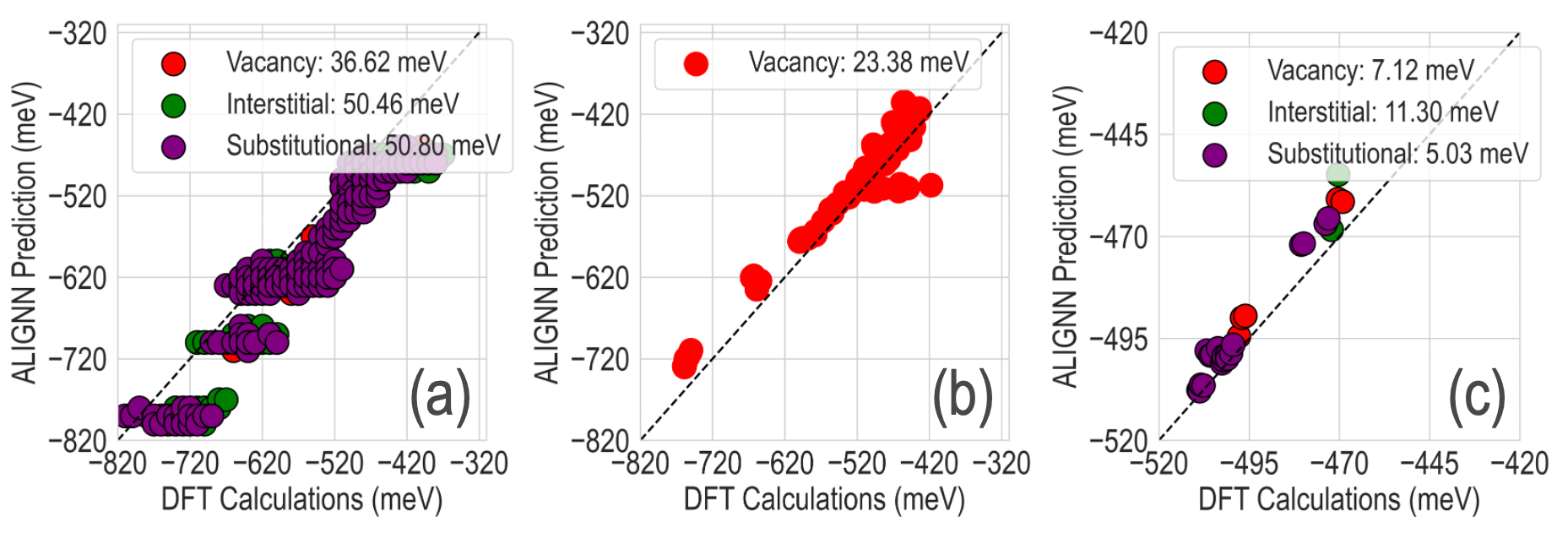}
    \caption{\label{fig:transferability} 
    Parity plots comparing ALIGNN predictions to DFT calculations under different training conditions: 
    (a) ALIGNN (trained solely on bulk structures) predictions vs. DFT calculations for the defect dataset.
    (b) ALIGNN (trained exclusively on interstitial defect structures) predictions vs. DFT calculations for the vacancy dataset. (c) ALIGNN (trained only on 2$\times$2$\times$2 supercell bulk and defect structures) predictions vs. DFT calculations for defects in a 3$\times$3$\times$3 supercell.}
\end{figure}

\section*{Graph neural network models for direct prediction of crystal formation energy}

ALIGNN was developed by Choudhary \textit{et al.}\cite{ALIGNN} and considers both two-body (bond lengths) and three-body interactions (bond angles). ALIGNN leverages both graph convolution layers and line graph convolution layers to capture short-range and long-range correlations in the crystal. For training the ALIGNN models, the learning rate was set to 0.001, an AdamW optimizer was used to update the weights and biases of the model, 4 graph convolution layers and 4 line graph layers were implemented, the cutoff radius was set to 6 Å with 12 nearest neighbors to create the crystal graph, and models were trained up to 90 epochs with a batch size of 8. We experimented with different training-validation-test splits of the dataset and found that the 60:20:20 ratio works the best. \\

ALIGNN models were trained to predict the CFE (PBE) from any given bulk or defect crystal structure, using only the neutral charge state structures at this stage. Parity plots capturing the performance of the optimized models are presented in \textbf{Figure \ref{fig:al_workflow}}, in terms of ALIGNN-predicted CFE vs DFT-computed CFE for only the test set data points. Models pictured in \textbf{Figure \ref{fig:al_workflow}(a-c)} are respectively trained only on bulk structures, only on defect structures, and on both bulk and defect structures; this distinction is made to understand how sensitive the models are to different types of configuration. As shown in \textbf{Figure \ref{fig:al_workflow}(a)}, the ALIGNN model for bulk structures alone shows a test prediction root mean squared error (RMSE) of 1.43 meV/atom, and this error remains practically unchanged for the combined model in \textbf{Figure \ref{fig:al_workflow}(c)}. For the defect-only ALIGNN model in \textbf{Figure \ref{fig:al_workflow}(b)}, test RMSE ranges from 3.09 meV/atom for interstitial defects to 4.87 meV/atom for substitutional defects to 8.36 meV/atom for vacancy defects. Each of these defect prediction errors comes down for the combined data model, proving the value of increasing the size and chemical and structural diversity of the training dataset.  \\

The training dataset contains a larger number of interstitial defects, followed by substitutional and vacancy defects: this is mostly a consequence of there being a lot more options for intersitial and substitutional defects in terms of extrinsic species from across the periodic table, and also the longer time it takes for DFT optimization of these defect structures, leading to more intermediate geometries. This results in the comparatively higher RMSE for CFE prediction of vacancy defects (6.47 meV/atom) than substitutional (3.84 meV/atom) and interstitial (2.04 meV/atom) defects. The ALIGNN predictions for all types of structures are highly accurate with vanishingly small errors considering the total range of CFE values. GNN models for defect predictions reported in the recent literature \cite{Mosquera2, Witman_2023, Rahman_Gollapalli} primarily focused on sampling defect configurations to train surrogate models, while we have adopted the approach of combining bulk and defect structures which enhances the overall generalizability and accuracy of the models. \\

\begin{figure}[!htbp]
\centering
\includegraphics[width=.9\linewidth]{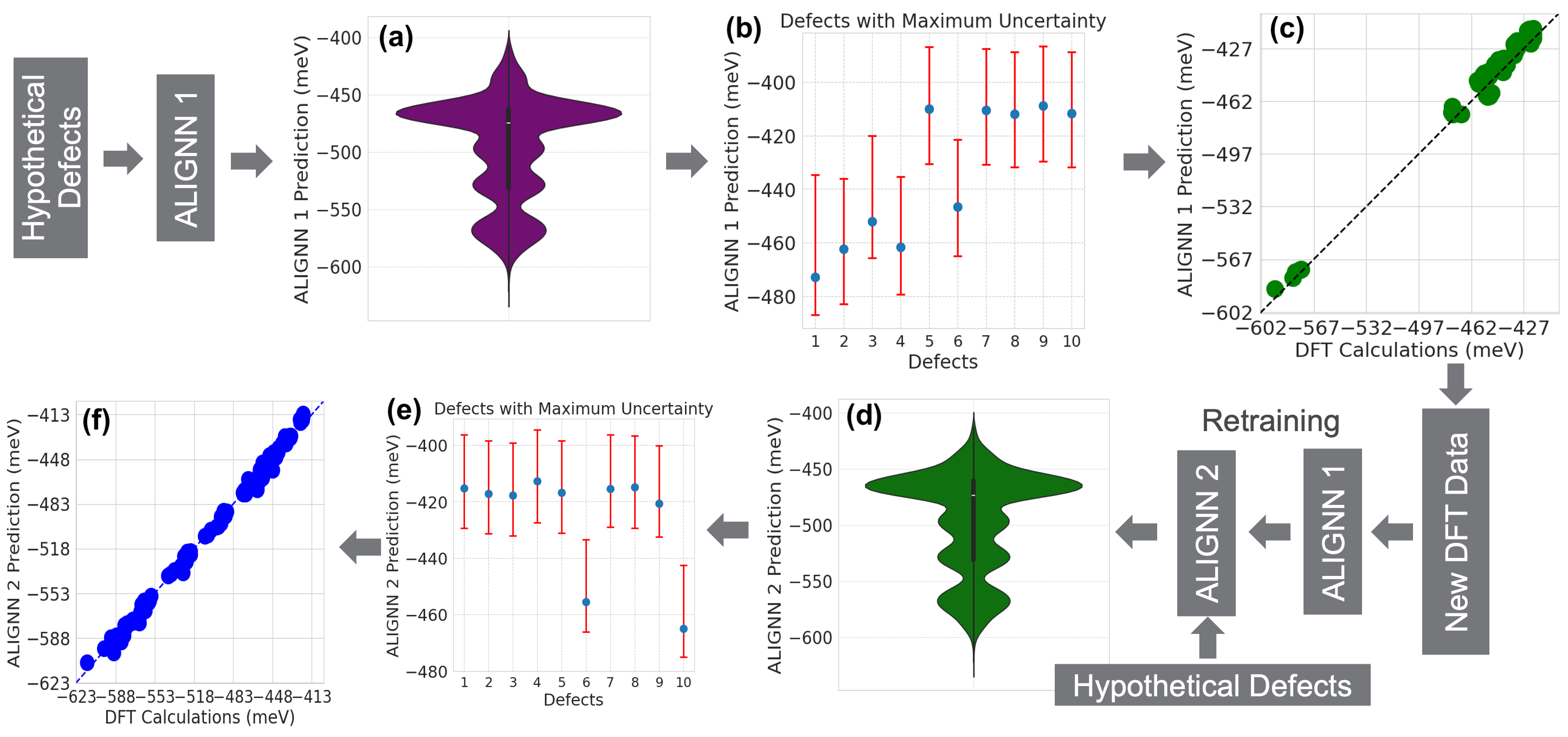}
\caption{\label{fig:al_cycle} (a) Violin plot showing the mean crystal formation energy (CFE) predicted by the initial ALIGNN-1 models, averaged across 100 models. (b) Defects with maximum uncertainties identified through ALIGNN-1 models. (c) Comparison of ALIGNN-1 predictions vs DFT calculations for 200 selected defects. (d) Violin plot of mean CFE from ALIGNN-2 models. (e) Defects with maximum uncertainties identified through ALIGNN-2 models. (f) Comparison of ALIGNN 2 predictions vs DFT calculations for the 200 selected defects.}
\end{figure}

Although state-of-the-art GNNs are very robust and powerful for modeling complex relationships within atomic structures \cite{Choudhary_DeCost, ALIGNN, Choudhary, MEGNET, CGCNN, Cheng_Dong_2021, Lee_Asahi, Fung}, their transferability beyond the trained dataset remains questionable. To evaluate this, we performed the following series of tests:

\begin{enumerate}
    \item An ALIGNN model was trained purely on bulk structures and then used to predict the CFE of defect structures.
    \item An ALIGNN model was trained only on interstitial defect structures and used to predict the CFE for vacancy and substitutional defects.
    \item An ALIGNN model trained exclusively on defects in 2$\times$2$\times$2 supercell structures and then used to predict the CFE of 3$\times$3$\times$3 supercell defect structures.
\end{enumerate}

ALIGNN shows poor transferability when trained only on bulk data and used to predict for defects; as shown in \textbf{Figure \ref{fig:transferability}(a)}, the prediction RMSEs range from $\sim$ 36 meV/atom for vacancies to $>$ 50 meV/atom for interstitial and substitutional defects. Since this model has not been exposed to specific configurations such as atomic relaxation around defect sites, it is unable to predict as accurately for defect structures as it does for bulk. \textbf{Figure~\ref{fig:transferability}(a)} also shows that ALIGNN uniformly under-predicts the CFE of all defect configurations, which could be attributed to the lower average CFE values in the bulk dataset compared to defects, as illustrated by the violin plot in \textbf{Figure \ref{fig:initial_pbe_dataset}(e)}. The model trained only on interstitial defects does a reasonable job for vacancy defects, but the RMSE values are larger than from the models in \textbf{Figure \ref{fig:al_workflow}} and there are some very clear outliers, as pictured in \textbf{Figure \ref{fig:transferability}(b)}. Lastly, when the model is trained on only 2$\times$2$\times$2 supercell structures and used to predict for 3$\times$3$\times$3 supercells, the predictions show good accuracy but a slight tendency to over-estimate the CFE, hinting at the fact that ALIGNN may be capable of extrapolating across supercell sizes. These results suggest that the GNN models for CFE prediction could generalize across types of chemistries, structures, and system sizes for particular cases, but in general may need to be retrained and fine-tuned for specific datasets. \\

\section*{Active Learning (AL) Workflow}

\begin{enumerate}[label=\Alph*.]
    \item \textbf{Training the Ensemble of ALIGNN Models:} We begin by training an ensemble of ALIGNN models to capture the variability and uncertainties associated with the predictions. To make this ensemble, we partition the original training set into multiple subsets, each containing a different combination of training, validation, and test data. A total of 100 different ALIGNN models have been trained, each on a unique subset of the data, allowing us to account for variability due to data partitioning. \textbf{Figure \ref{fig:al_workflow}(b)} illustrates the ALIGNN predictions on the test dataset (we name it ALIGNN-1) vs. DFT calculations from 100 different ALIGNN models, highlighting the standard deviation in the predictions along with the mean.

    \item \textbf{Prediction Across Expanded Defect Chemical Space:} After training the ensemble of ALIGNN-1 models, we utilize them to predict the CFE of all the defects in the expanded chemical space. For each configuration, predictions are made using all 100 models in the ensemble, yielding a distribution of predictions. This approach enables us to not only obtain the mean prediction but also to quantify the uncertainty associated with each prediction. \textbf{Figure \ref{fig:al_cycle}(a)} shows the violin plot of predicted mean CFE (averaged across 100 ALIGNN models) made across the entire set of defects.  

    \item \textbf{Uncertainty Quantification:} The uncertainty of each prediction is quantified by analyzing the standard deviation of CFE among the predictions made by the 100 ALIGNN-1 models. In our AL framework, we employed the maximum uncertainty (MU) acquisition function \cite{Farache_Verduzco}. The MU criterion is defined as $\text{MU}(x) =  \sigma(x)$, where $\sigma(x)$ denotes the standard deviation (uncertainty) of the prediction. \textbf{Figure \ref{fig:al_cycle}(b)} highlights the defects that maximize the MU acquisition function identified by the ALIGNN-1 models.

    \item \textbf{Active Learning via Bayesian optimization and New DFT Calculations:} We utilized Bayesian optimization to refine the predictions of the ALIGNN-1 models by selecting the 200 configurations that maximize the chosen acquisition function, prioritizing the most informative data points. These selected configurations were then used to launch new DFT calculations, ensuring that the model iteratively improves its accuracy and predictive performance. Although the model initially has not encountered certain defects, such as those in the Cd$_{x}$Zn$_{1-x}$Te composition, the predictions from ALIGNN-1 models are reasonable. \textbf{Figure \ref{fig:al_cycle}(c)} shows ALIGNN-1 prediction (energy of the initial input defect structure, \textit{viz.}, unoptimized energy) vs. DFT calculation (unoptimized energy) for 200 selected defects based on acquisition function. To further improve the performance of ALIGNN-1 models, these new calculations are then incorporated into the training set, iteratively improving the accuracy of the model.

    \item \textbf{Final Model Performance:} Following the first iteration of the AL loop, we retrain ALIGNN-1 models and get new ALIGNN models (we name it ALIGNN-2) which are used to make predictions across the remaining defect space. The violin plot of predicted mean CFE (averaged across 100 ALIGNN-2 models) made on the remaining defect configurations is presented in \textbf{Figure \ref{fig:al_cycle}(d)}. The predictions are again evaluated using the acquisition function as shown in \textbf{Figure \ref{fig:al_cycle}(e)}, and 200 new configurations that maximize its value are selected for new DFT calculations. We observe a significant improvement in the ALIGNN predictions as depicted in \textbf{Figure \ref{fig:al_cycle}(f)}, closely matching the DFT-computed CFE across the selected defects. This rapid convergence after just one training cycle highlights the effectiveness of the AL approach in enhancing model performance, even in underexplored regions of the chemical space. Later on, newly obtained DFT data is again incorporated into the training set and we retrained the model (ALIGNN-3). Finally, the ALIGNN-3 models are used to predict the remaining unexplored defects.The rationale behind choosing the 200 defects that maximizes the MU acquisition function is driven by a careful consideration of our computational budget and capabilities.
\end{enumerate}

\section*{Detailed Description of ALIGNN-Based Geometry Optimization}

\begin{figure}[!htbp]
\centering
\includegraphics[width=.9\linewidth]{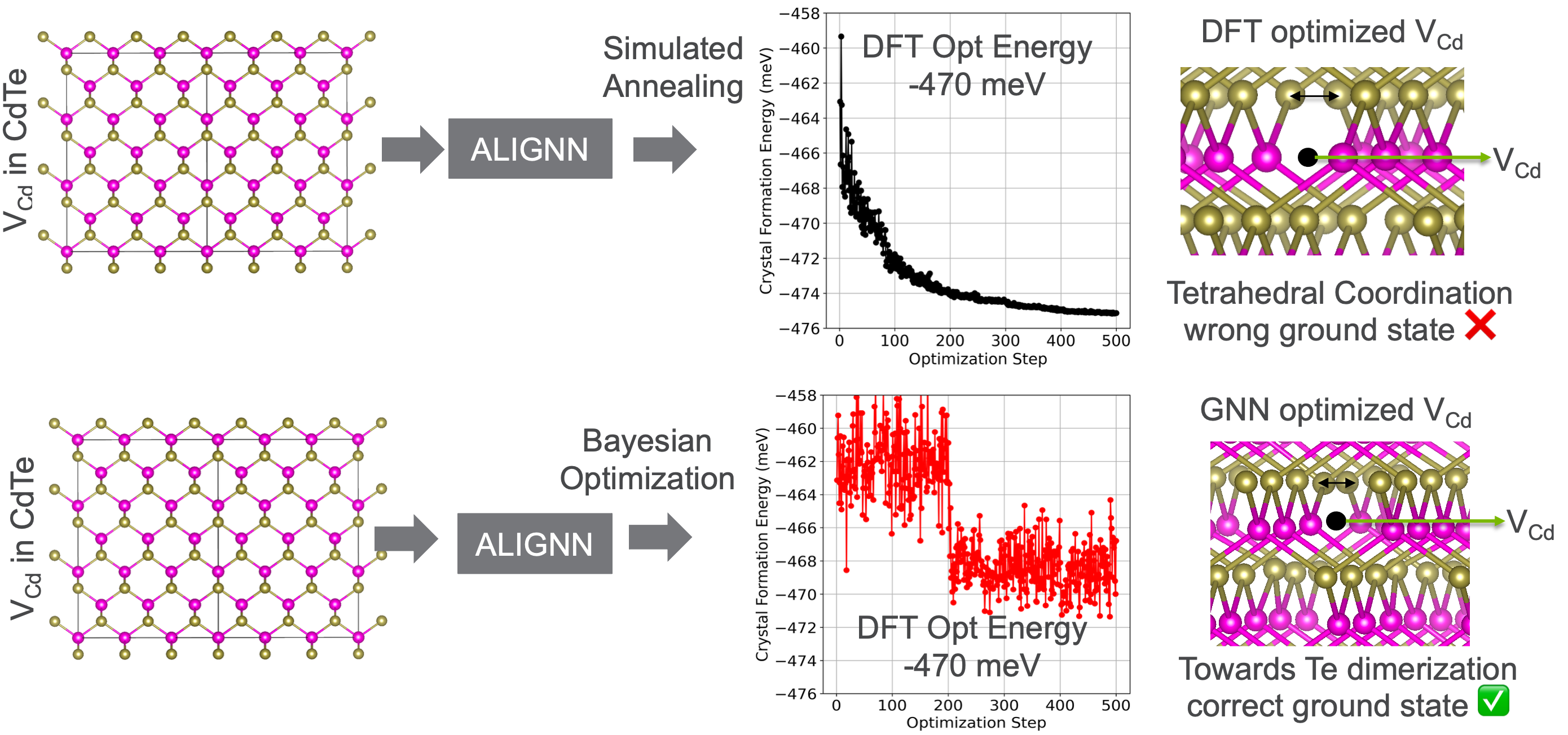}
\caption{\label{fig:alignn_optimization_v_cd} 
Optimization of Cd vacancy (V$_{Cd}$) in CdTe using the trained ALIGNN model with two different optimization strategies: simulated annealing and Bayesian optimization.
}
\end{figure}

\begin{figure}[!htbp]
\centering
\includegraphics[width=.9\linewidth]{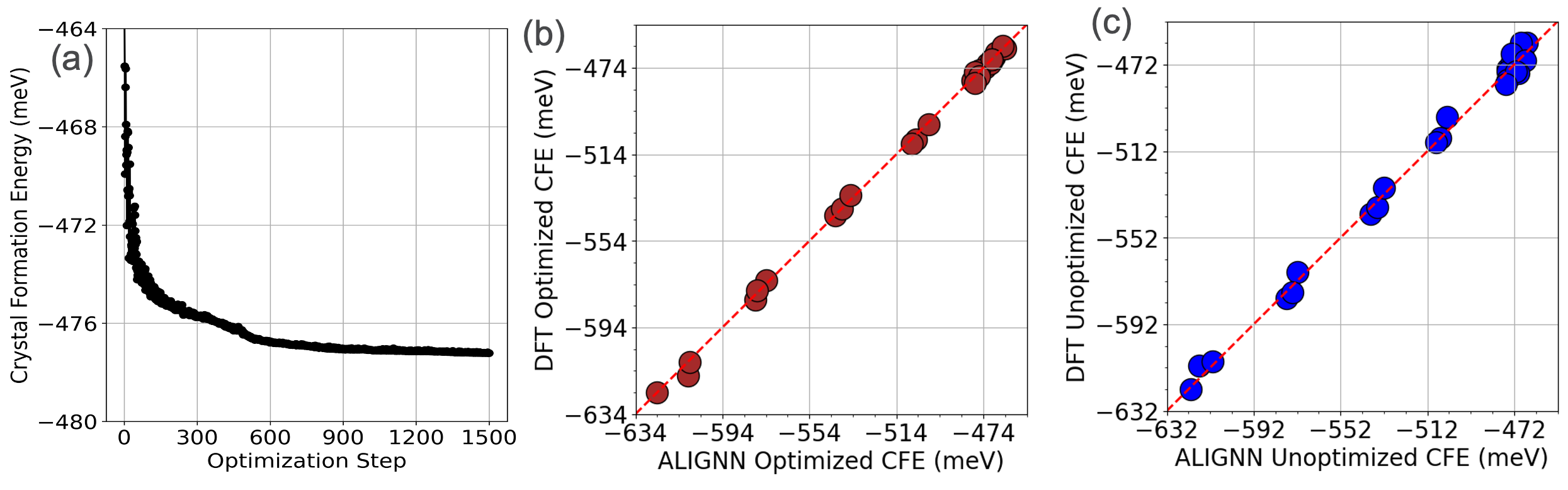}
\caption{\label{fig:alignn_optimization} a) Simulated annealing optimization of a Cl$_{Te}$ defect in CdSe$_{0.50}$Te$_{0.50}$ using ALIGNN-predicted crystal formation energy (CFE), showing energy minimization over successive steps.  
(b) Parity plot comparing ALIGNN-optimized CFE with DFT-optimized CFE across a set of defect structures, demonstrating strong agreement.  
(c) Parity plot comparing ALIGNN-unoptimized CFE with DFT-unoptimized CFE. }
\end{figure}

\FloatBarrier

\begin{figure}[!htbp]
\centering
\includegraphics[width=.9\linewidth]{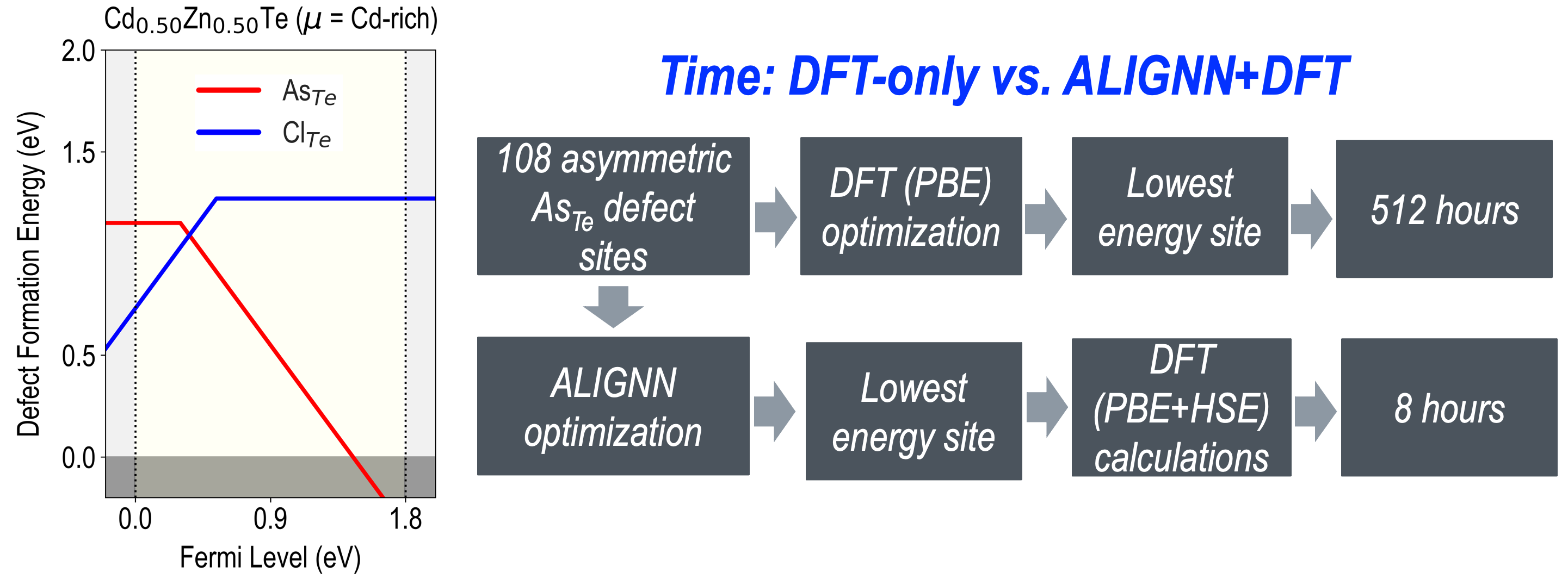}
\caption{\label{fig:alignn_cdznte} 
Comparison of computational efficiency between a DFT-only workflow and the ALIGNN+SA+DFT approach for evaluating Cl$_{Te}$ and As$_{Te}$ defects in Cd$_{0.50}$Zn$_{0.50}$Te under Cd-rich conditions.  Left: Defect formation energies (DFEs) as a function of Fermi level, computed using the HSE06+SOC functional on top of the PBE optimized structures.   Right: Workflow comparison showing that direct DFT relaxation of all 108 symmetry-inequivalent As$_{Te}$ configurations requires approximately 512 hours, while the ALIGNN+SA model identifies the lowest-energy site in minutes, followed by ALIGNN+PBE optimization. A single static HSE06+SOC calculation on the ALIGNN+PBE optimized configuration reduces the total computational time to just 8 hours.
}
\end{figure}

Here, we combined the ALIGNN predictions with two commonly used optimization techniques to achieve energy minimization of new defect configurations: 
\begin{enumerate}
    \item Simulated Annealing (SA) \cite{Cheng_Gong}, which is a gradient-free stochastic method employing random atomic perturbations guided by an annealing schedule.
    \item Bayesian Optimization (BO) \cite{Cheng_Gong}, which uses Gaussian processes to iteratively identify low energy configurations through atomic displacement exploration. 
\end{enumerate}

As an illustrative example, both simulated annealing and Bayesian optimization were applied to optimize a V$_{Cd}$ defect within a 3$\times$3$\times$3 CdTe supercell \textbf{(see Figure \ref{fig:alignn_optimization_v_cd})}. Each method rapidly identified stable defect structures within minutes, significantly faster than traditional DFT methods. Simulated annealing notably discovered a Te–Te dimer configuration, which standard DFT often overlooks without prior chemical intuition.  \\

Coupling ALIGNN prediction of CFE values with SA or BO enabled gradient-free energy minimization via atomic displacements applied within a cut-off radius around the defect center. These algorithms systematically searched for the lowest energy configuration using ALIGNN predictions at each step, allowing fast and guided optimization. As a case study, we applied both algorithms to optimize a Cd vacancy (V$_{Cd}$) defect in a 3$\times$3$\times$3 CdTe supercell, and the results are pictured in \textbf{Figure ~\ref{fig:alignn_optimization_v_cd}}. Either method reaches a general energy convergence within 300 to 400 optimization steps which take a total of a few minutes to complete, but SA performs much better than BO: not only does it find a lower energy structure, but it actually discovers the configuration featuring a Te–Te dimer, which was reported by Kavanagh et al. \cite{sean_1} and which is easily missed by standard DFT optimization. To further evaluate the optimization capability of the ALIGNN model, we combined it with SA to optimize a variety of defects across the CdSe$_{x}$Te$_{1-x}$ and Cd$_{x}$Zn$_{1-x}$Te chemical space as shown in \textbf{Figure ~\ref{fig:alignn_optimization}}. For example, Cl$_{Te}$ in Cd$_{0.50}$Zn$_{0.50}$Te was successfully optimized using ALIGNN+SA, as shown in \textbf{Figure ~\ref{fig:alignn_optimization} (a)}.  \\

\begin{table}[htbp]
\centering
\caption{Performance of pretrained machine-learning force-field (MLFF) models applied directly to the PBE test dataset without retraining. The reported errors correspond to the root-mean-square error (RMSE) in crystal formation energy (CFE).}
\label{tab:pretrained_mlff_comparison}
\begin{tabular}{l l c}
\hline
\textbf{Model} & \textbf{Pretrained Version} & \textbf{RMSE (meV/atom)} \\
\hline
MACE \cite{mace} 
& \texttt{MACE\_MPtrj\_2022.9.model} 
& 93.79 \\
M3GNet \cite{m3gnet} 
& \texttt{M3GNet-MatPES-PBE-v2025.1-PES} 
& 62.10 \\
CHGNet \cite{chgnet} 
& \texttt{CHGNet-MatPES-PBE-2025.2.10-2.7M-PES}
& 59.50 \\
\hline
\end{tabular}
\end{table}

\textbf{Figure~\ref{fig:alignn_cdznte} }illustrates the DFEs of As$_{Te}$ and Cl$_{Te}$ in Cd$_{0.50}$Zn$_{0.50}$Te under Cd-rich conditions as a function of E$_{F}$, computed using the HSE06+SOC functional on top of PBE-optimized lowest-energy sites. The right panel compares two workflows and highlights the computational advantage of incorporating ALIGNN+SA-based optimization. In the conventional DFT-only approach, all 108 symmetry-inequivalent As$_{Te}$ configurations must be relaxed individually to identify the lowest-energy structure, requiring approximately 512 hours. In contrast, the ALIGNN+DFT workflow first uses the trained ALIGNN+SA model to rapidly evaluate and optimize the CFE for all configurations, identifying the most stable site in minutes. DFT calculations (PBE followed by HSE06+SOC) are then performed on the ALIGNN-predicted lowest-energy structure, reducing the total computational cost to just 8 hours. The next section presents an attempt to improve upon this by moving towards gradient-based optimization using force fields.

\section*{PBE Machine learning Force Field (MLFF) Model}

M3GNet-based \cite{m3gnet} MLFF models were trained using the energies, atomic forces, and stresses extracted from GGA–PBE calculation trajectories. Radial and three-body cutoffs were set to 6~\AA\ and 6~\AA, respectively. The loss was a weighted sum of RMSE for energies, forces, and stresses (weights = 1, 1, and 0.01, respectively). Training was performed on an NVIDIA A100 (80~GB) with batch size 64 and learning rate $5\times10^{-4}$ until convergence. Geometry optimization with the MLFF used the FIRE algorithm in ASE~\cite{ase_1,ase_2} with convergence criteria of mean atomic force $<10^{-5}$~eV/\AA\ or a maximum of 100 ionic steps. Models were trained separately for structures in five different charge states. To improve model accuracy on difficult configurations where predictions were poor, we used a two-stage training process:

\begin{itemize}
  \item \textbf{Warm-up:} In the first stage, we trained the model for a small number of epochs using uniform sampling. This helps the model develop a basic understanding of the data.

  \item \textbf{Error-aware reweighting:} Next, we used the warm-up model to predict energies and forces for all training samples. Based on the prediction errors, we assigned a score to each sample. Samples with larger errors were considered harder. These error scores were then converted into sampling weights. Limits were applied to avoid extremely large or small weights and cap any outliers. These weights were passed to a \texttt{WeightedRandomSampler} in PyTorch, which increased the likelihood of selecting harder examples during training. The validation set remained unweighted. The reweighting step was repeated every 10–20 epochs to keep the weights up to date. This method helps the model focus more on difficult configurations while still learning broadly, which leads to better performance on complex regions of the data.
\end{itemize}

\begin{figure*}[t]
\centering
\includegraphics[width=.9\linewidth]{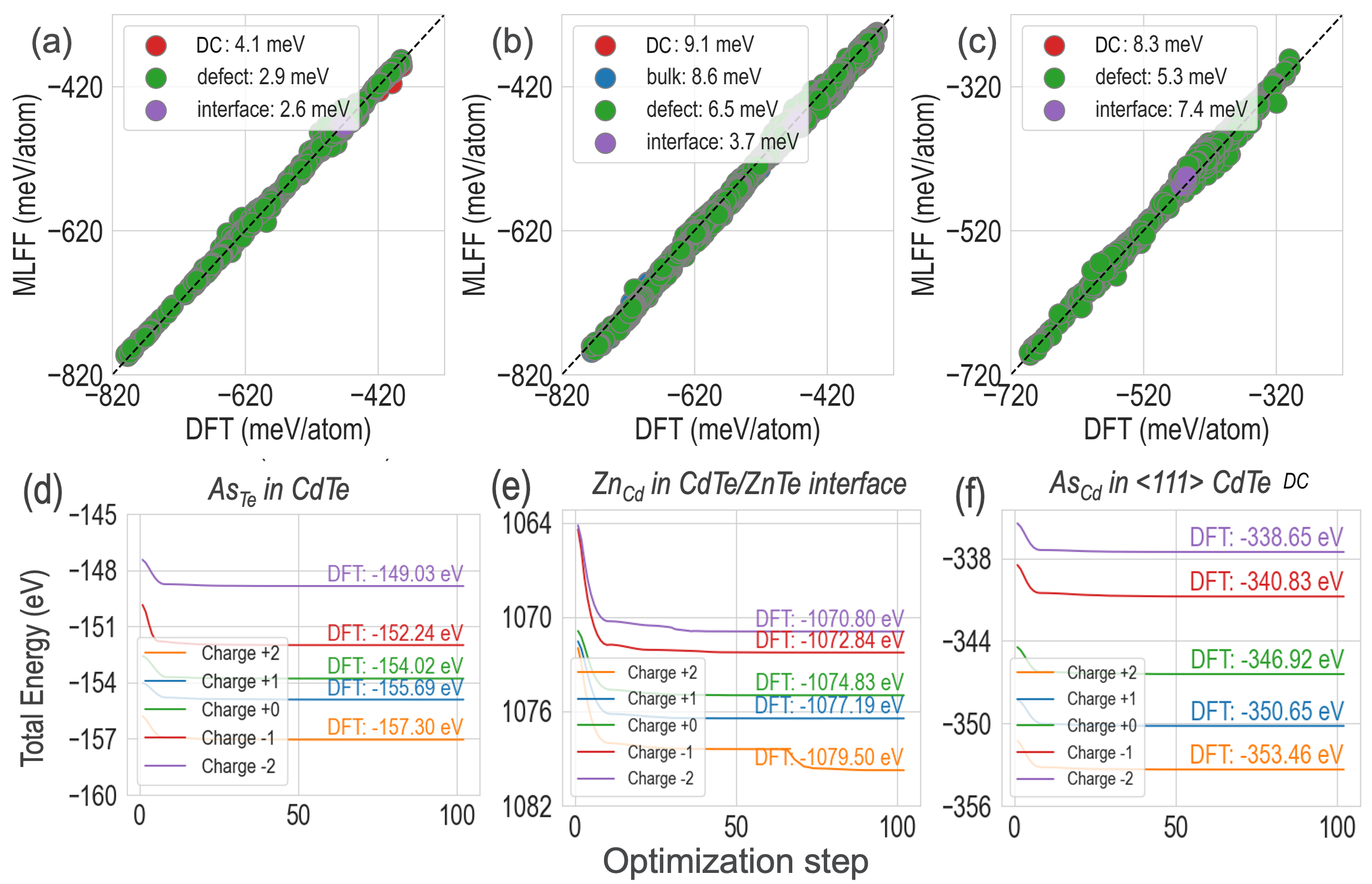}
\caption{\label{fig:gga_m3gnet_model_1}
Performance of M3GNet-MLFF models trained on the GGA-PBE dataset, shown in terms of predicted vs DFT crystal formation energy parity plots. The models were trained separately for (a) defect configurations with charge \textit{q=+1}, (b) neutral \textit{q=0} defect and bulk configurations, and (c) defect configurations with charge \textit{q=-1}. Here, "bulk" refers to pristine supercells without any defects, "defect" means bulk supercells containing a point defect or defect complex, "interface" corresponds to defects located at CdTe-ZnTe interfaces, and "DC" indicates defects situated in CdTe dislocation core structures. Plots in (d–f) show the geometry optimization process taking into account 5 charge states for different example defect configurations: (d) a As$_{Te}$ defect in CdTe, (e) a Zn$_{Cd}$ defect at a CdTe/ZnTe interface, and (f) As$_{Cd}$ in a <111> CdTe  dislocation core structure (DC).}
\end{figure*}

\begin{figure}[!htbp]
  \centering
  \includegraphics[width=0.8\columnwidth]{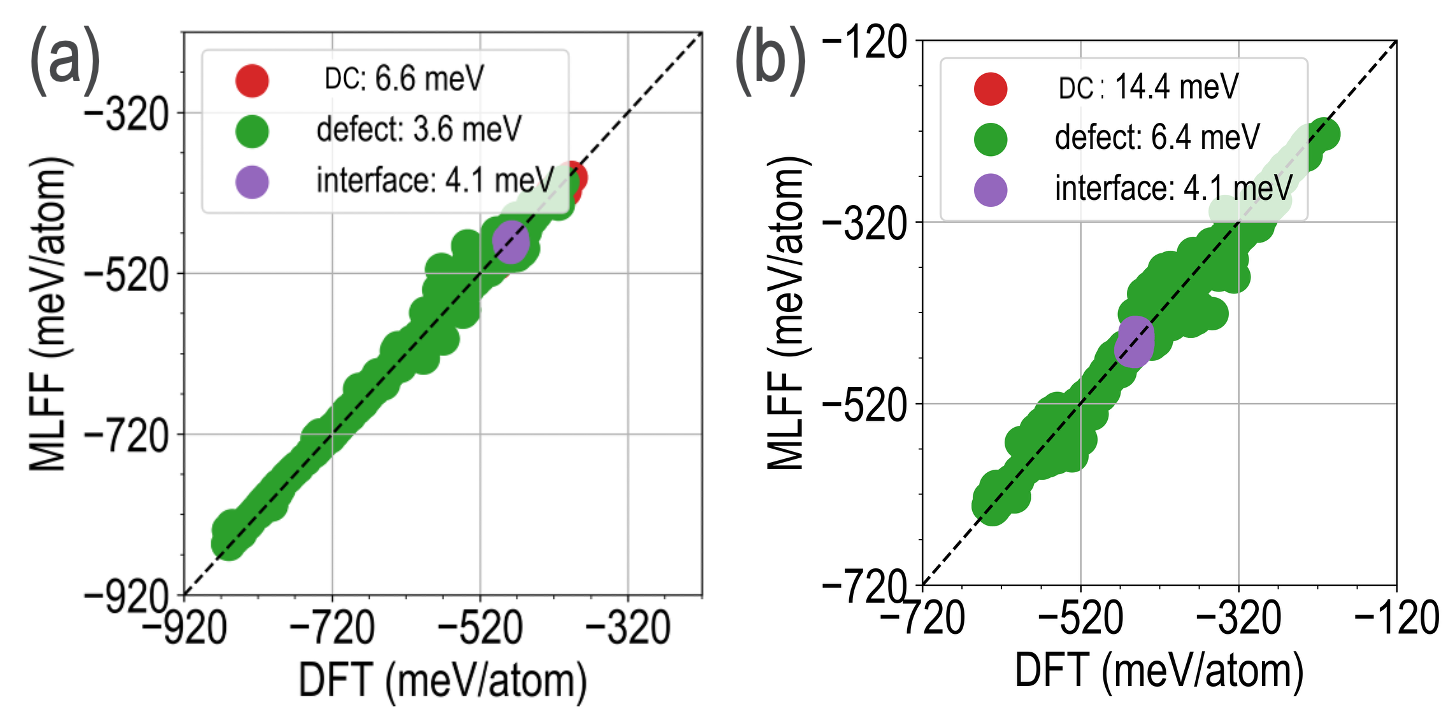}
  \caption{ Parity plots for M3GNet-MLFF models trained on the GGA dataset, shown in terms of predicted vs actual (from DFT) crystal formation energies, trained separately for (a) defect configurations with charge \textit{q=+2}, (b) defect configurations with charge \textit{q=-2}. Here, "defect" represents bulk supercells containing a point defect or defect complex, "interface" corresponds to defects located at CdTe-ZnTe interfaces, and "DC" indicates defects situated at CdTe dislocation core.}
  \label{fig:gga_m3gnet_model_2}
\end{figure}

\textbf{Figure~\ref{fig:gga_m3gnet_model_1}(a--c)} show parity plots for the M3GNet-MLFF models trained for charge states \(q = +1\), \(q = 0\), and \(q = -1\). Models for the \(q=+2\) and \(q=-2\) charge states are presented in \textbf{Figure~\ref{fig:gga_m3gnet_model_2}}. Each parity plot compares DFT-computed CFE for test set points with the corresponding values from the MLFF prediction for different types of structures: bulk (pristine supercells without defects), defects (bulk supercells containing a single point defect or defect complex), defects at CdTe-ZnTe interfaces, and defects in CdTe dislocation core (DC). Overall, the MLFF predictions show remarkably low RMSE values for all types of bulk and defect configurations, similar to the ALIGNN models. \\

For \(q = 0\), the test prediction RMSE for bulk structures is 8.6 meV/atom, a similarly low value of 6.8 meV/atom for defect structures, and 3.7 meV/atom for interface. Dislocation core defect structures show slightly larger errors of $>$ 9.1 meV/atom, which is expected given their more complex local environments and atomic rearrangements. Similar errors were seen for charged defect structures as well. The overall agreement between DFT and MLFF predictions remains generally strong, indicating the robustness of the model across diverse chemical and structural environments. \\

\begin{figure}[!htbp]
\centering
\includegraphics[width=.9\linewidth]{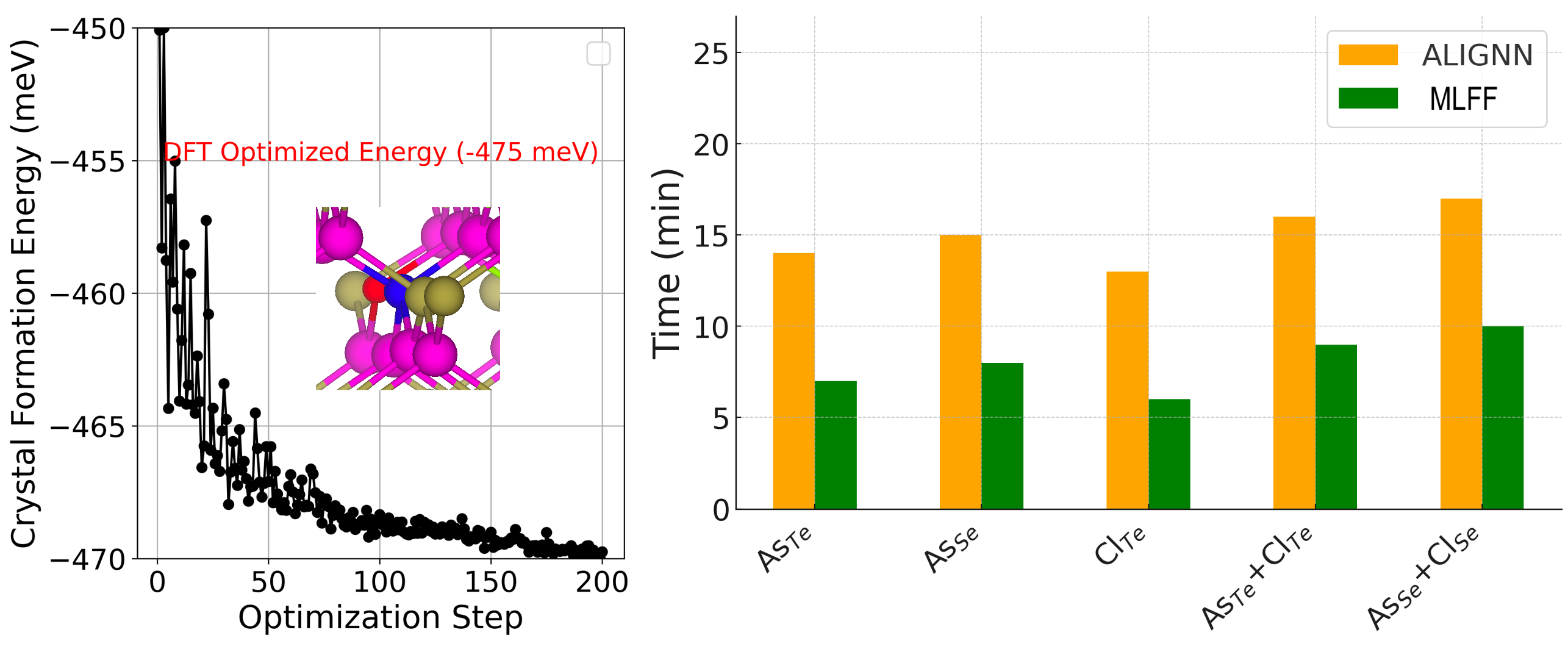}
\caption{\label{fig:alignn_vs_m3gnet}
(a) Optimization of an As$_{Te}$ + Cl$_{Te}$ complex in CdSe$_{0.12}$Te$_{0.88}$ using an M3GNET-MLFF model, and (b) comparison of the time taken by ALIGNN and the MLFF for optimizing selected defects in CdSe$_{0.12}$Te$_{0.88}$.
}
\end{figure}

One of the major advantages of having an MLFF model rather than a direct energy prediction model is the ability to use predicted atomic forces to perform geometry optimization based on gradient-based energy minimization, which is more computationally efficient than gradient-free optimization as shown in \textbf{Figure~\ref{fig:alignn_vs_m3gnet}}. For example, we optimized the As$_{Te}$+Cl$_{Te}$ defect in CdSe$_{0.12}$Te$_{0.88}$ using both ALIGNN (direct energy) and M3GNet (MLFF) and found that M3GNet achieved the optimization at a substantially lower computational cost compared to ALIGNN. \textbf{Figure~\ref{fig:gga_m3gnet_model_1}(d--f)} show three different examples of using the MLFF model for optimizing challenging defect configurations: an As$_{Te}$ substitutional defect in CdTe, a Zn$_{Cd}$ defect in a CdTe-ZnTe interface structure, and an As$_{Cd}$ substitutional defect in a CdTe dislocation core structure. These cases represent chemically and structurally complex environments that are often found in devices with polycrystalline semiconductor thin films. The optimized \(q = +2, +1, 0, -1, -2\)  MLFF models were able to successfully capture local atomic rearrangements and produce low energy configurations consistent with DFT benchmarks. In each case, the MLFF-optimized configuration energy matches well with the DFT-optimized energy. Energy minimization is achieved in approximately 100 steps, with the entire relaxation process completing within a few minutes. \\

\begin{figure*}[t]
\centering
\includegraphics[width=.9\linewidth]{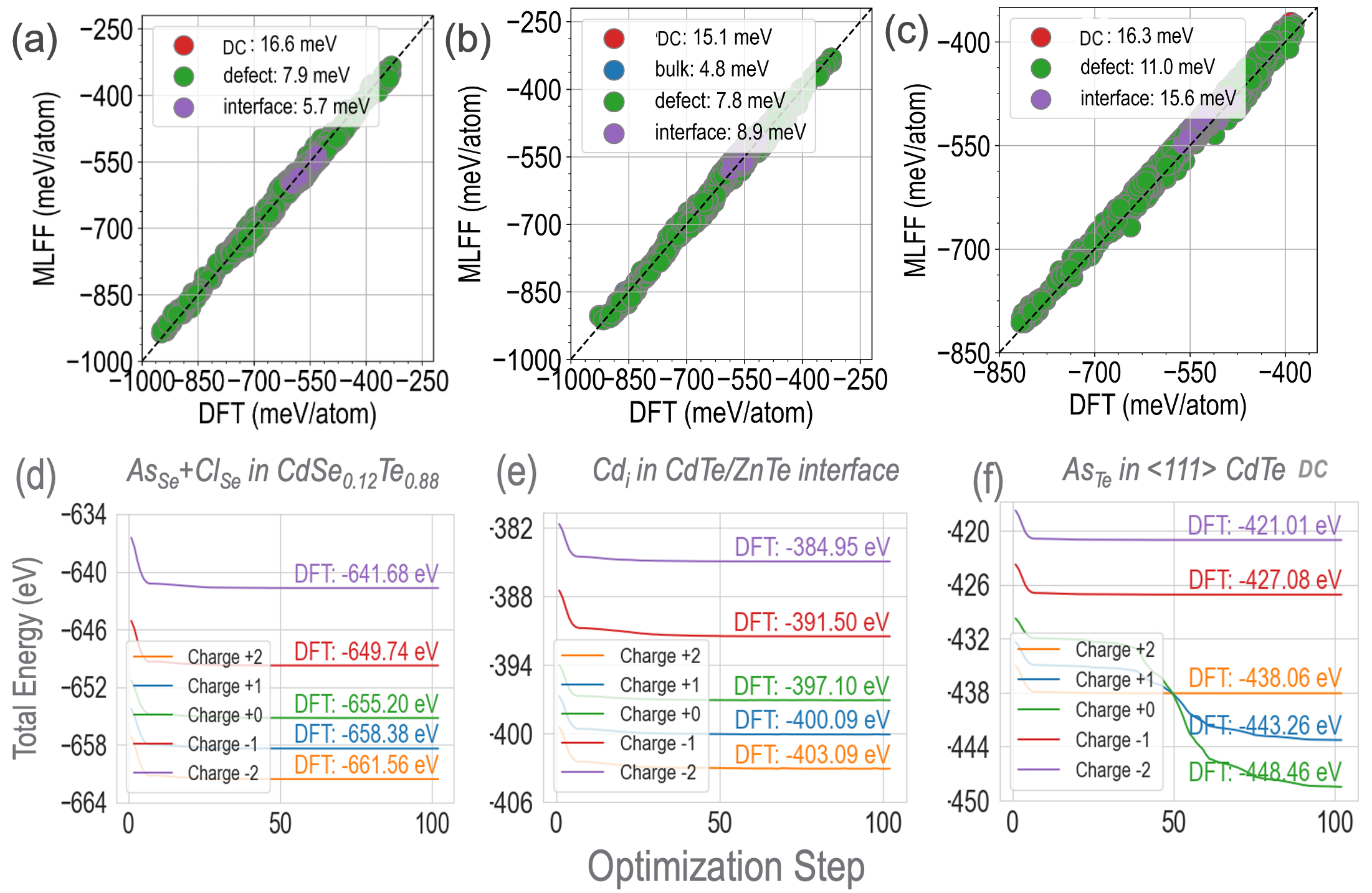}
\caption{\label{fig:hse_m3gnet_model_old_1}
Performance of the M3GNet-MLFF models trained on the augmented HSE06 dataset, shown in terms of predicted vs DFT crystal formation energy parity plots. The models were trained separately for (a) defect configurations with charge \textit{q=+1}, (b) neutral \textit{q=0} defect and bulk configurations, and (c) defect configurations with charge \textit{q=-1}. Here, "bulk" refers to pristine supercells without any defects, "defect" means bulk supercells containing a point defect or defect complex, "interface" corresponds to defects located at CdTe-ZnTe interfaces, and "DC" indicates defects situated at CdTe dislocation core. Plots in (d–f) show the geometry optimization process taking into account 5 charge states for different example defect configurations: (d) an As$_{Se}$+Cl$_{Se}$ defect complex in CdSe$_{0.12}$Te$_{0.88}$, (e) a Cd$_{i}$ defect at the CdTe/ZnTe interface, and (f) As$_{Te}$ in <111> CdTe  dislocation core structure.}
\end{figure*}

\begin{figure}[!htbp]
  \centering
  \includegraphics[width=0.8\columnwidth]{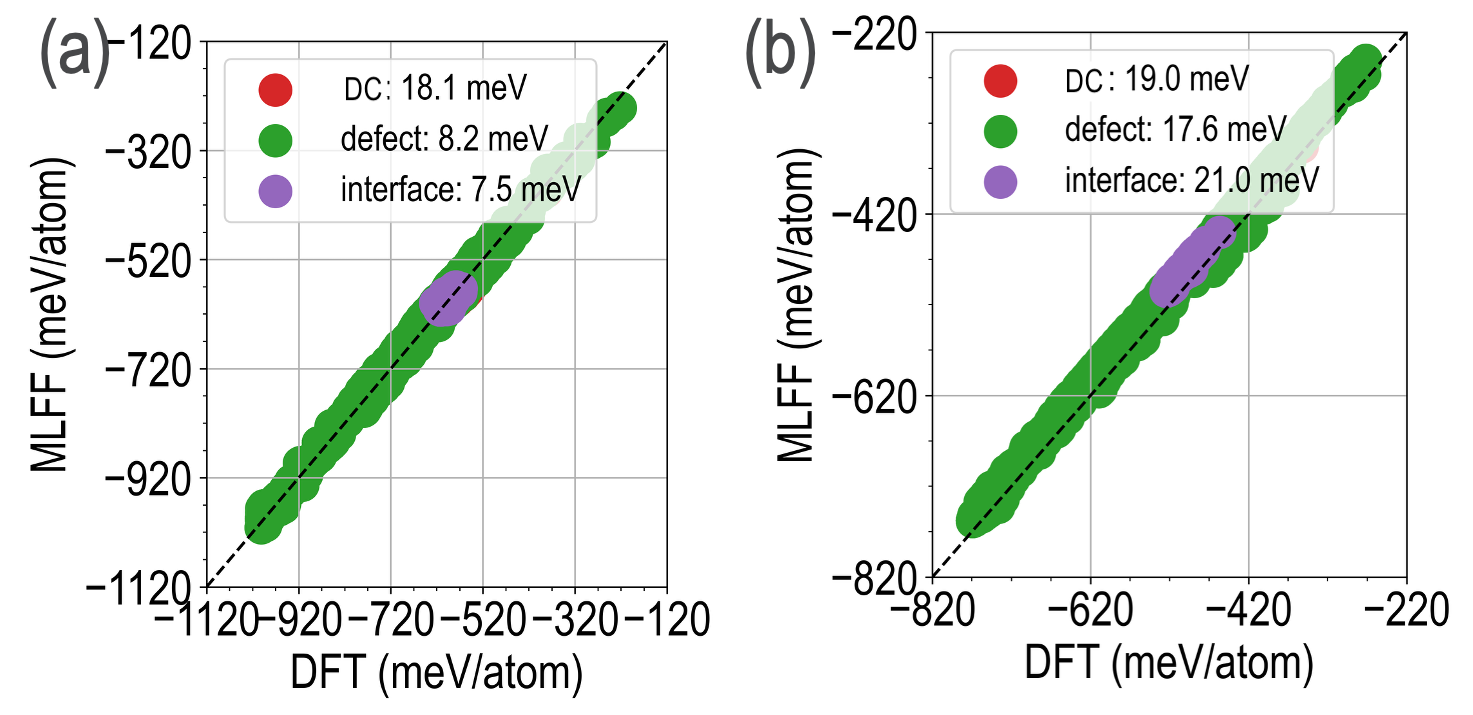}
  \caption{ Parity plots for M3GNet-MLFF models trained on the HSE06 dataset, shown in terms of predicted vs actual (from DFT) crystal formation energies, trained separately for (a) defect configurations with charge \textit{q=+2}, (b) defect configurations with charge \textit{q=-2}. Here, "defect" represents bulk supercells containing a point defect or defect complex, "interface" corresponds to defects located at CdTe-ZnTe interfaces, and "DC" indicates defects situated at CdTe dislocation core structure.}
  \label{fig:hse_m3gnet_model_old_2}
\end{figure}

\FloatBarrier

\begin{figure}[!htbp]
    \centering
    \includegraphics[width=0.4\textwidth]{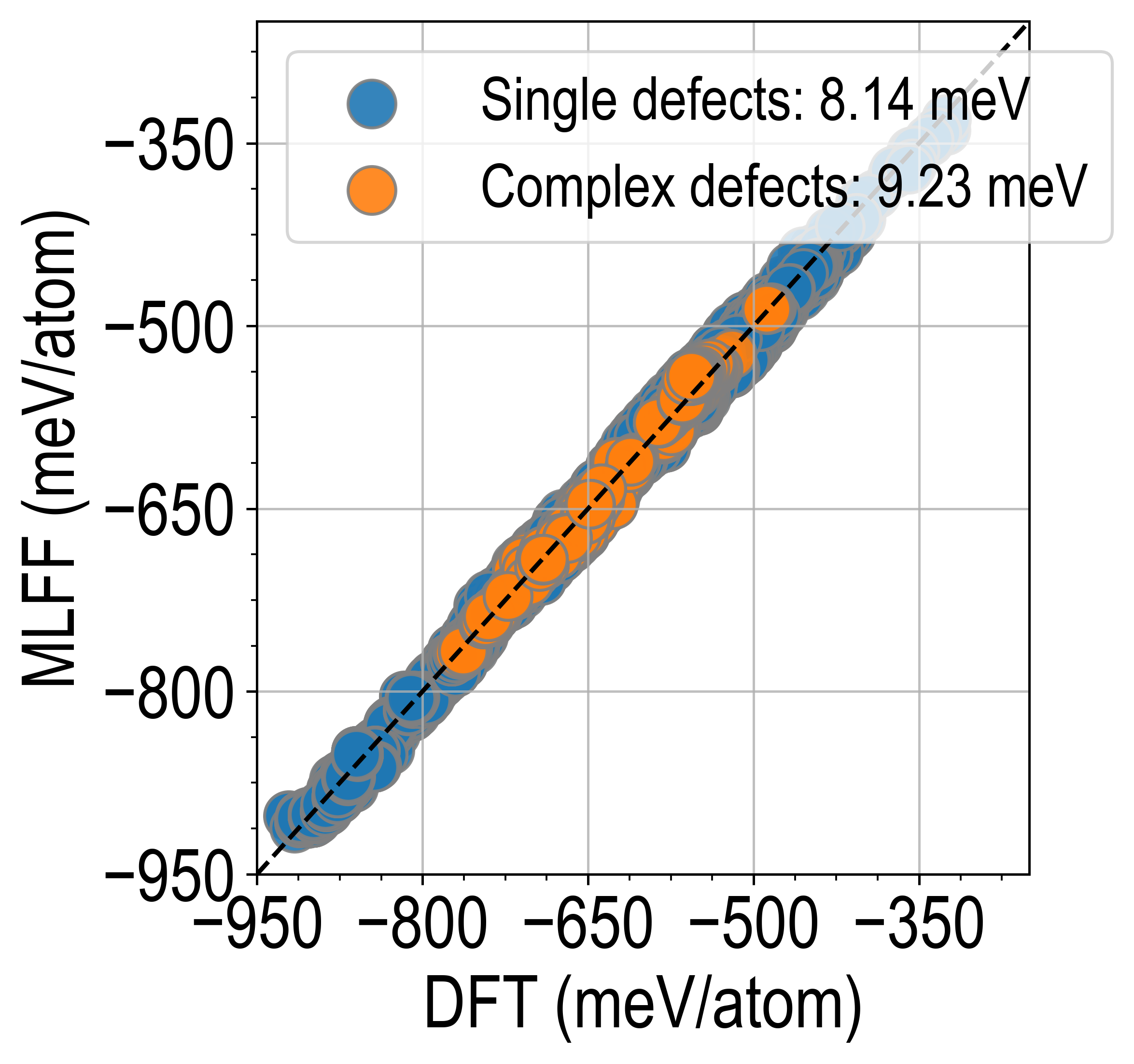}
    \caption{
        Parity plot comparing DFT and MLFF-predicted crystal formation energies (CFE) for single and complex defects (\textit{q}=0 charge state). }
    \label{fig:single_vs_complex}
\end{figure}

\FloatBarrier

\begin{figure*}[t]
  \centering
  \includegraphics[width=1\textwidth]{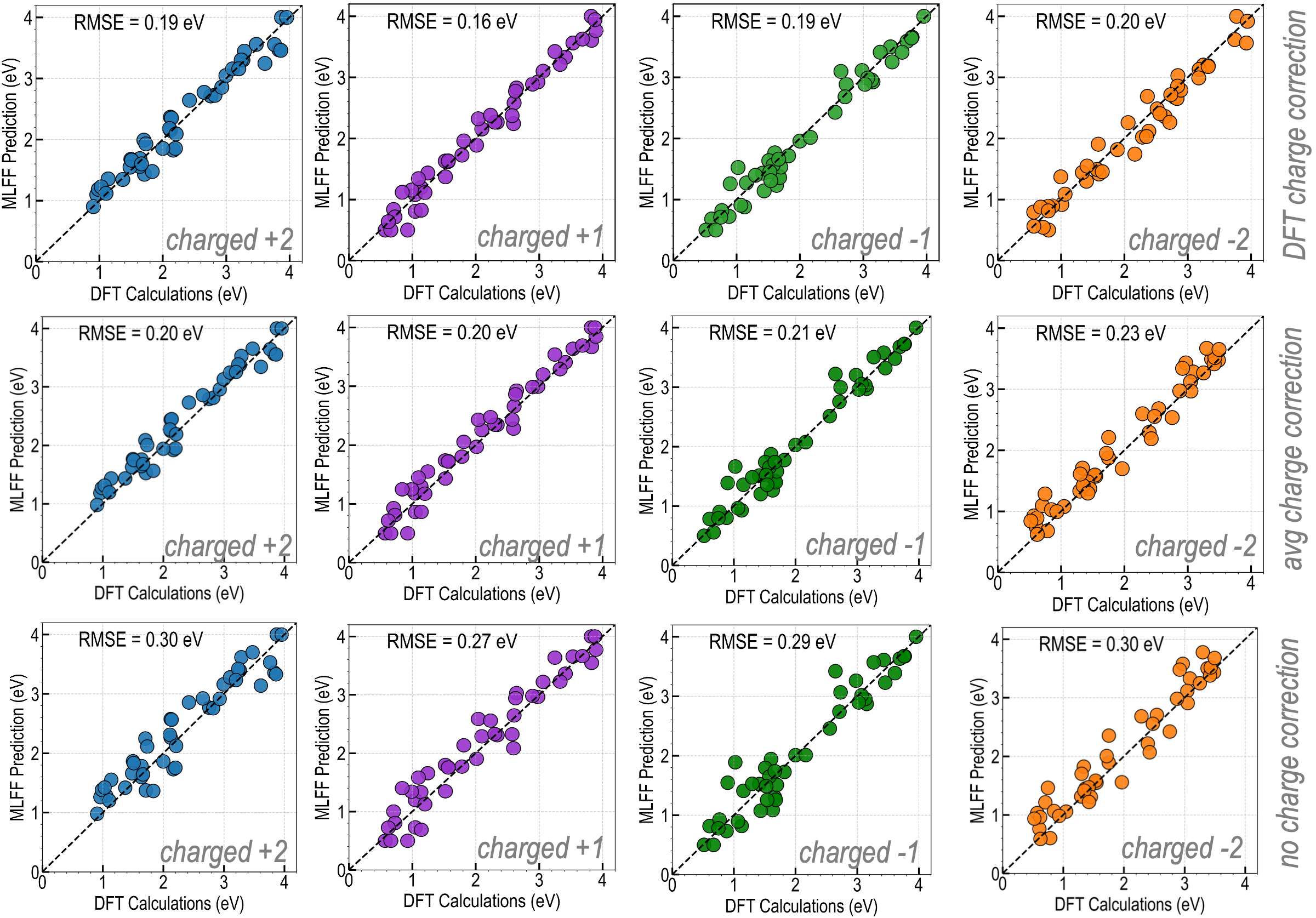}%  <-- changed
\caption{Parity plots comparing MLFF-predicted defect formation energies with DFT values for different charge states and charge–correction schemes. Each panel shows the MLFF prediction plotted against the corresponding DFT value for charged defects with $q = +2$, $+1$, $-1$, and $-2$.  The first row uses actual DFT charge corrections, the second row uses averaged charge corrections for each charge, and the third row applies no charge correction.}
  \label{fig:charge_correction}
\end{figure*}

\begin{figure*}[t]
  \centering
  \includegraphics[width=0.6\textwidth]{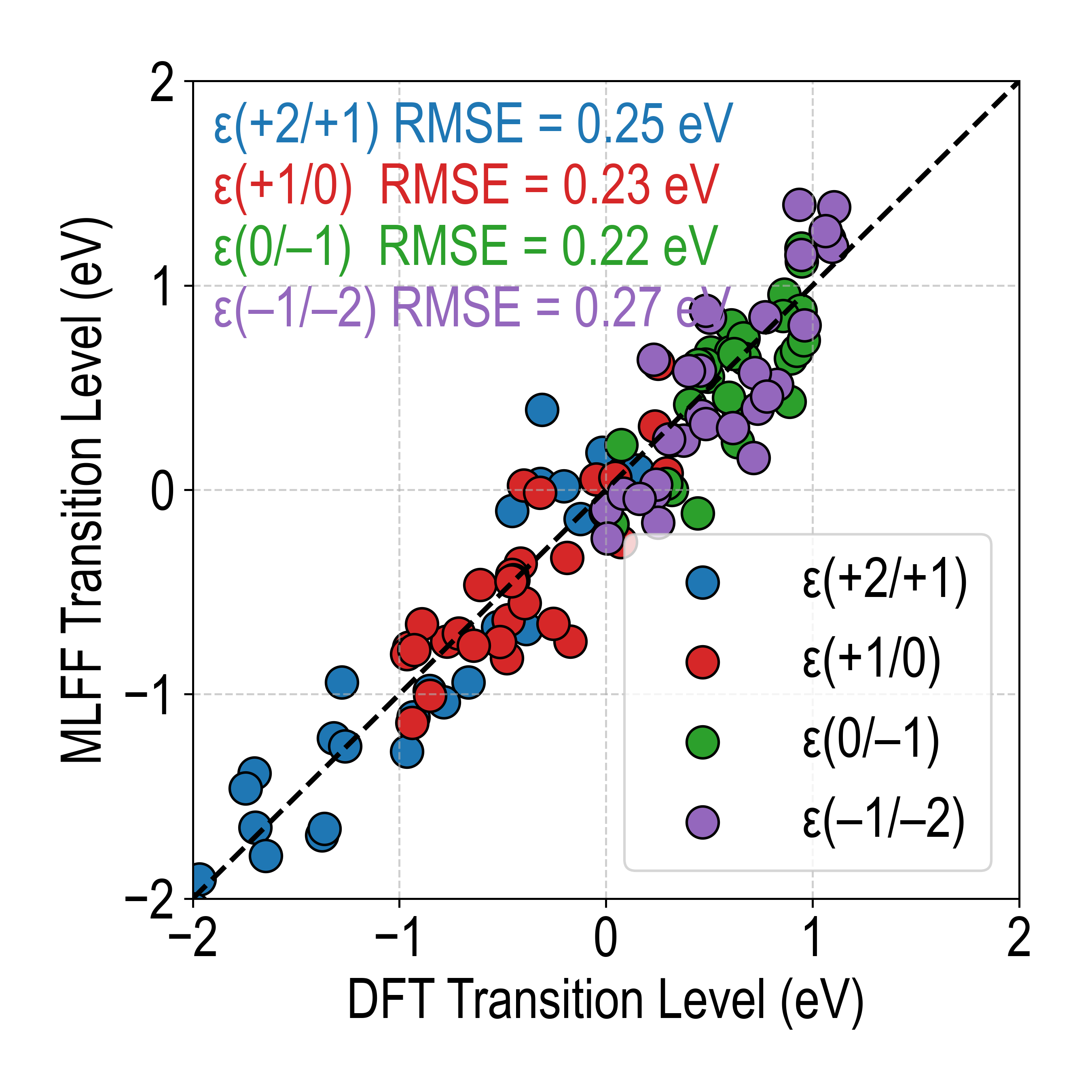}%  <-- changed
\caption{Parity plot comparing MLFF-predicted and DFT-calculated charge-transition levels, with average charge correction value applied to MLFF prediction in each charge state. Four types of transition levels are calculated, namely +2/+1, +1/0, 0/-1, and -1/-2.}
  \label{fig:transition_level_plot}
\end{figure*}

\begin{figure*}[t]
  \centering
  \includegraphics[width=0.4\textwidth]{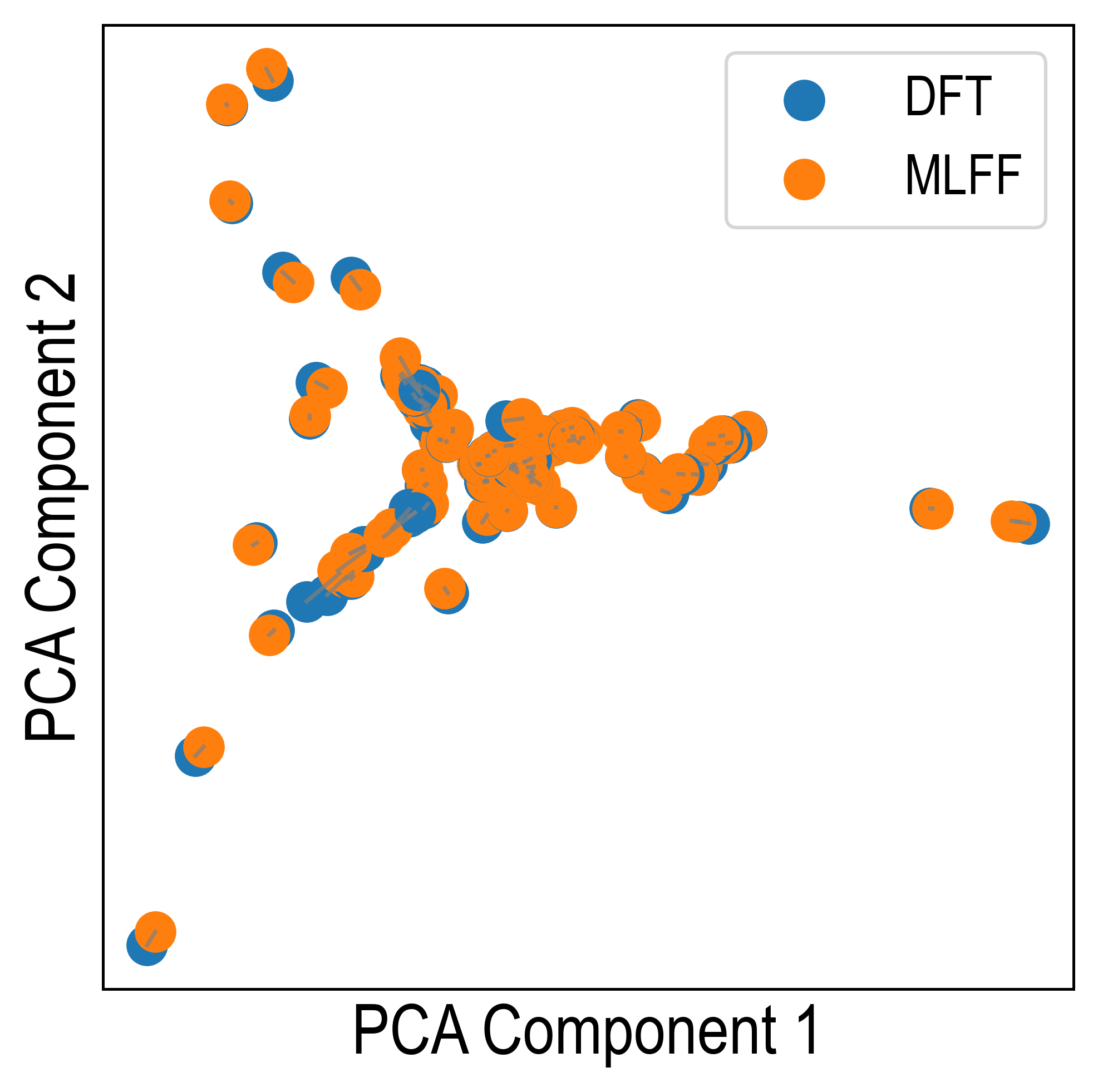}%  <-- changed
\caption{Comparison of DFT- and MLFF-relaxed structures using PCA on SOAP descriptors \cite{dscribe_1, dscribe_2}. Each point represents the structural fingerprint of a relaxed configuration projected onto a two-dimensional PCA space. Blue and orange markers correspond to DFT and MLFF optimizations, respectively. The near-overlapping distribution of DFT and MLFF points demonstrates that the latter accurately reproduces DFT-level structural relaxations across diverse bulk and defect configurations.}
  \label{fig:pca}
\end{figure*}

\begin{figure*}[t]
  \centering
  \includegraphics[width=\textwidth]{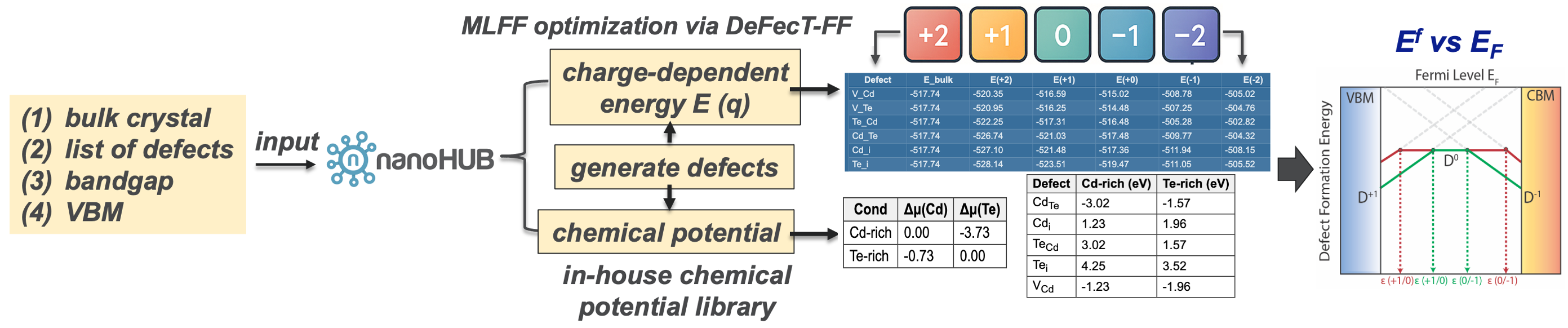}%  <-- changed
  \caption{The \texttt{DeFecT-FF} tool takes as input the bulk crystal, list of defects, bandgap, and VBM, performs MLFF-based geometry optimization, leverages an in-house chemical potential library to calculate charge-dependent energies $E(q)$ and defect formation energies $E_f$ as a function of the Fermi level $E_F$, and finally constructs $E_f$–$E_F$ diagrams for defect thermodynamics. This tool is accessible via a nanoHUB web application.}
  \label{fig:defectff_nanoHUB}
\end{figure*}

\begin{figure}[!htbp]
\centering
\includegraphics[width=.9\linewidth]{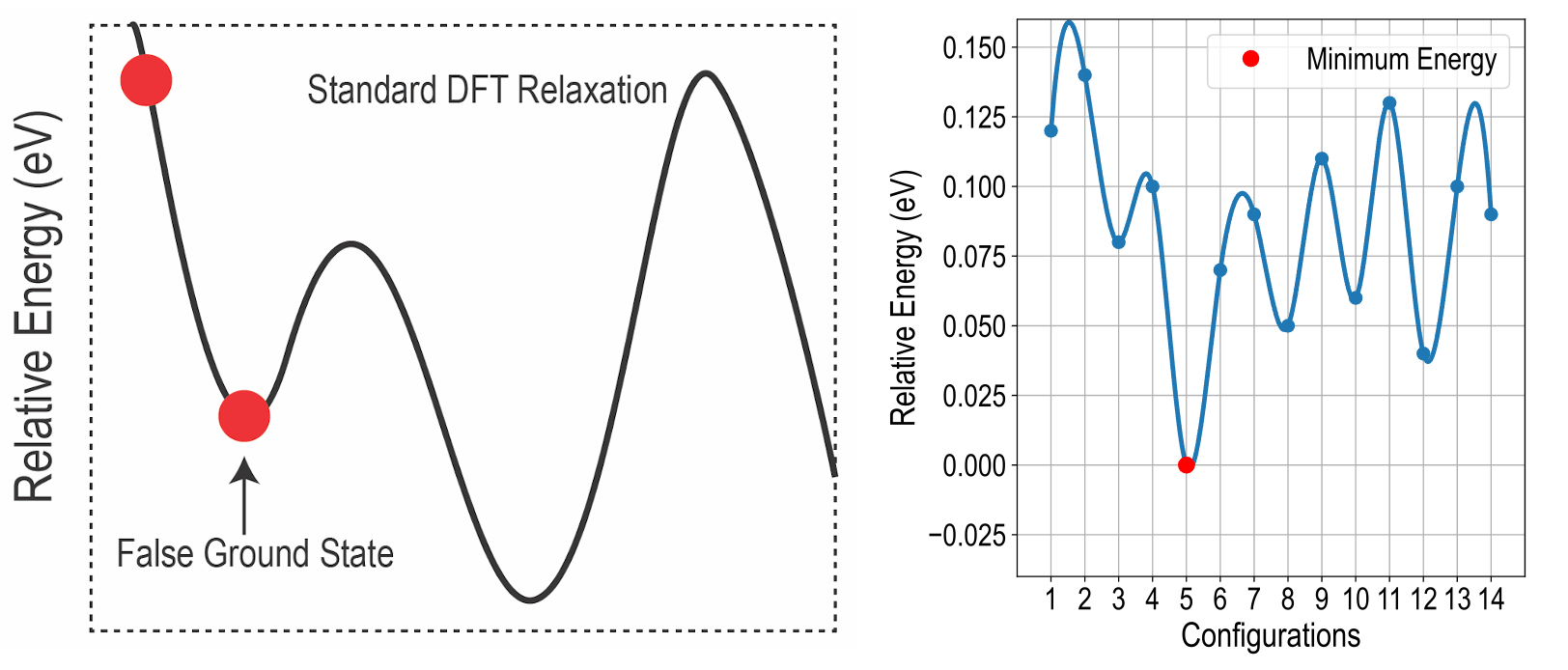}
\caption{\label{fig:snb_pes} Illustration of the configuration search for As$_{\mathrm{Se}}$ defects. The left panel shows how a standard DFT relaxation can become trapped in a false ground state, while the right panel presents the relative energies of 14 symmetry-broken configurations obtained using the HSE-based MLFF. The true lowest-energy configuration (configuration 5) is highlighted in red. }
\end{figure}

\FloatBarrier

\begin{table}[h!]
\centering
\caption{Comparison of RMSE (eV) in MLFF defect formation energies for three correction schemes:
(i) no charge correction,
(ii) average charge-offset correction, and
(iii) DFT charge correction applied to MLFF predictions.}
\begin{tabular}{lccc}
\hline
\textbf{Charge State} &
\textbf{No Correction} &
\textbf{Average Offset Correction} &
\textbf{DFT Charge Correction} \\
\hline
$q = -2$ & 0.30 & 0.23 & 0.20 \\
$q = -1$ & 0.29 & 0.21 & 0.19 \\
$q = 0$  & 0.16       & 0.16       & 0.16 \\
$q = +1$ & 0.27 & 0.20 & 0.16 \\
$q = +2$ & 0.30 & 0.20 & 0.19 \\
\hline
\end{tabular}
\label{tab:rmse_charge_corrections}
\end{table}

\begin{table}[h!]
\centering
\caption{RMSE for crystal formation energy prediction broken down by defect type for the neutral ($q=0$) charge state.}
\label{tab:rmse_by_defect_type}
\begin{tabular}{l c}
\hline\hline
Defect Type & RMSE (meV/atom) \\
\hline
Vacancy        & 9.5  \\
Antisite       & 8.1  \\
Substitutional & 8.6  \\
Interstitial   & 7.6  \\
\hline\hline
\end{tabular}
\end{table}

\begin{table}[h!]
\centering
\caption{Composition-resolved test RMSE for neutral crystal formation energy (CFE) prediction across bulk and defect configurations in all compounds in the DeFecT-FF chemical space. Binary compounds show slightly lower errors compared to ternary alloys due to reduced compositional disorder.}
\label{table:rmse_by_composition}
\begin{tabular}{l c}
\hline\hline
Compound & RMSE (meV/atom) \\
\hline
CdTe                         & 7.2  \\
CdSe                         & 7.5  \\
ZnTe                         & 7.8  \\
CdSe$_{0.25}$Te$_{0.75}$     & 8.6  \\
CdSe$_{0.50}$Te$_{0.50}$     & 9.1  \\
CdSe$_{0.75}$Te$_{0.25}$     & 9.4  \\
Cd$_{0.25}$Zn$_{0.75}$Te     & 8.9  \\
Cd$_{0.50}$Zn$_{0.50}$Te     & 9.3  \\
Cd$_{0.75}$Zn$_{0.25}$Te     & 8.4  \\
\hline\hline
\end{tabular}
\end{table}

\begin{table}[h!]
\centering
\caption{Out-of-distribution (OOD) and in-distribution (ID) test results for crystal formation energy prediction. OOD RMSE corresponds to compositions not included in the MLFF training set, while ID RMSE corresponds to the same compositions after being added to the training set, demonstrating the improvement in predictive accuracy upon inclusion.}
\label{tab:ood_rmse}
\begin{tabular}{l c c}
\hline\hline
Composition & OOD RMSE (meV/atom) & ID RMSE (meV/atom) \\
\hline
CdSe$_{0.12}$Te$_{0.88}$ & 12.4 & 8.3 \\
CdSe$_{0.06}$Te$_{0.94}$ & 12.8 & 8.7 \\
\hline\hline
\end{tabular}
\end{table}

\section*{Charge Correction Using the Freysoldt Scheme}

Charged defects in periodic boundary conditions introduce spurious electrostatic interactions between the charged defect, its periodic images, and the compensating background charge. To correct for these finite-size effects, we employed the Freysoldt~\cite{Freysoldt2014-en} correction scheme as implemented in the \texttt{sxdefectalign} code. The correction energy consists of two components: an image-charge term ($E_{\mathrm{PC}}$) and a potential alignment term ($q\,\Delta V$):

\begin{equation}
E_{\mathrm{corr}} = E_{\mathrm{PC}} + q\,\Delta V.
\end{equation}

The image-charge correction is estimated using an isotropic model charge distribution screened by the static dielectric constant $\varepsilon$ of the host material:

\begin{equation}
E_{\mathrm{PC}} = \frac{q^{2}\alpha}{2\,\varepsilon\,L},
\end{equation}

where $q$ is the defect charge state, $\alpha$ is the Madelung constant for the supercell geometry, and $L$ is the effective lattice parameter of the supercell. The potential alignment term $\Delta V$ is obtained from the planar-averaged electrostatic potential difference between the defective and bulk supercells in the far-field region:

\begin{equation}
\Delta V = V_{\mathrm{defect}}^{\mathrm{far}} - V_{\mathrm{bulk}}.
\end{equation}

\end{document}